# Large-Area Intercalated 2D-Pb/Graphene Heterostructure as a Platform for Generating Spin-Orbit Torque

Alexander Vera[1,2], Boyang Zheng[3,4], Wilson Yanez[2,3], Kaijie Yang[2,3], Seong Yeoul Kim[5], Xinglu Wang[5], Jimmy C. Kotsakidis[6], Hesham El-Sherif[7], Gopi Krishnan[7], Roland J. Koch[8], Timothy Bowen[1,2], Chengye Dong[4,9], Yuanxi Wang[4], Maxwell Wetherington[10], Eli Rotenberg[8], Nabil Bassim[7,11], Adam L. Friedman[6], Robert M. Wallace[5], Chaoxing Liu[2,3], Nitin Samarth[2,3,4,9], Vincent H. Crespi[1,2,3,4,9,10,12], Joshua A. Robinson[1,2,3,4,9,10,12,13*]

1. Department of Materials Science and Engineering, The Pennsylvania State University, University Park, PA, USA.
2. Center for Nanoscale Science, The Pennsylvania State University, University Park, PA, USA.
3. Department of Physics, The Pennsylvania State University, University Park, PA, USA.
4. 2-Dimensional Crystal Consortium, The Pennsylvania State University, University Park, PA, USA.
5. Department of Materials Science and Engineering, The University of Texas at Dallas, Dallas, TX, USA.
6. Laboratory for Physical Sciences, College Park, MD, USA.
7. Department of Materials Science and Engineering, McMaster University, Hamilton, Ontario, Canada
8. Advanced Light Source, Lawrence Berkeley National Laboratory, Berkeley, CA, USA
9. Center for 2-Dimensional and Layered Materials, The Pennsylvania State University, University Park, PA, USA
10. Materials Research Institute, The Pennsylvania State University, University Park, PA, USA
11. Canadian Centre for Electron Microscopy, McMaster University, Hamilton, Ontario, Canada
12. Department of Chemistry, The Pennsylvania State University, University Park, PA, USA.
13. Center for Atomically Thin Multifunctional Coatings, The Pennsylvania State University, University Park, PA, USA

* jrobinson@psu.edu

## Abstract

**A scalable platform to synthesize ultrathin heavy metals may enable high efficiency charge-to-spin conversion for next-generation spintronics. Here we report the synthesis of air-stable, epitaxially registered monolayer Pb underneath graphene on SiC (0001) by confinement heteroepitaxy (CHet). Diffraction, spectroscopy, and microscopy reveal CHet-based Pb intercalation predominantly exhibits a mottled hexagonal superstructure due to an ordered network of Frenkel-Kontorova-like domain walls. The system's air stability enables *ex-situ* spin torque ferromagnetic resonance (ST-FMR) measurements that demonstrate charge-to-**

**spin conversion in graphene/Pb/ferromagnet heterostructures with a 1.5× increase in the effective field ratio compared to control samples.**

## Introduction

Advances in next-generation technologies demand materials design that can utilize extraordinary properties to a practical macro-scale. One such property domain is spintronics, which seeks to exploit the spin degree of freedom to encode information with less volatility, lower power consumption, and increase speed compared to conventional charge-based semiconductor devices.[1,2] The integration of spintronics into current solid-state technology requires efficient charge-to-spin conversion in, for example a current-induced spin polarization (CISP) layer underneath a ferromagnet.[3] This approach depends on materials with strong spin-orbit coupling (SOC), such as ultrathin heavy metals[4–6] or graphene with induced SOC.[7–10] Thus a promising material platform for spintronics is epitaxial graphene (EG) on silicon carbide (SiC).[11,12] Auspiciously, EG on SiC may be decoupled from the substrate by intercalation[13] of various atomic species[14–16] at elevated temperatures, including p-block metals,[17–23] rare-earth metals,[24–31] transition metals,[32–39] and alkali/alkaline-earth metals,[40–45] along with compounds[46–48] and alloys,[49,50] providing a flexible approach to not only generate and statically tune the properties of quasi-freestanding EG, but also to create a range of ambient-stable quasi-2D crystals sandwiched between EG and SiC. The observation of emergent spin-based phenomena, such as Rashba SOC, due to Sn[51] and Au[52,53] intercalation, and Pb intercalation on graphene/Pt(111),[54] further promote the potential of intercalation in spin-selective technologies.

Ultrathin Pb films on Si(111) exhibit large spin-orbit induced gaps[55–57] and enhanced Zeeman-protected type-II superconductivity.[58,59] In addition, recent predictions anticipate Pb on SiC(0001) to host a non-trivial antivortex spin texture.[60] Thus, intercalation of Pb in EG/SiC is an attractive candidate for *ex-situ* spintronics studies. Previous reports suggest a complex set of distinct phases of Pb after intercalation; namely, monolayer Pb(110) and Pb(111) showing a striped and hexagonal (10×10) Moiré periodicity[22,61–65] and charge-neutral QFEG,[22,64,65] a twisted honeycomb plumbene structure rotated ±7.5° from graphene showing 1D edge states,[66] monolayer Pb quasi-(1×1) to SiC(0001) with a periodic domain boundary network,[67] and a lower density amorphous phase.[68] Here, we demonstrate micron-scale intercalation of 2D-Pb through confinement heteroepitaxy (CHet),[18] a recently developed methodology that has been shown to improve the intercalation efficacy of lighter p-block metals through monolayer EG. CHet based Ga intercalation (2D-Ga) induces an increase in graphene's surface potential[69] and a set of ultra-low frequency (ULF) peaks between 26 – 119 $cm^{-1}$;[70] these characteristic features provide a facile method of coverage identification that is extendable to the Pb case, where we find two distinct contrasts in electron microscopy which correlate with Raman spectroscopy mapping of Pb-ULF peaks and Pb 4f core level photoemission counts. Low energy electron diffraction (LEED) and scanning tunneling microscopy (STM) elucidate a 10×10 superstructure; however, high-resolution scanning transmission electron microscopy (STEM) and experimental band structure measurements are consistent with a 1×1 arrangement. We explain this discrepancy through a detailed first-principles model where CHet-based EG/Pb/SiC relaxes compressive stress in the 1×1 arrangement through the formation of Frenkel-Kontorova-like (FK) domains within monolayer Pb that are separated by vacancy line defects, similar to results from Schädlich et al.[67] Using micron-scale coverage analysis, we attempt a spin-torque ferromagnetic resonance (ST-FMR) device using an optimized EG/2D-Pb as a CISP layer, finding a 1.5× increase in the effective field ratio over hydrogenated control samples. In doing so, we demonstrate intercalation of Pb in EG/SiC as a promising material platform for *ex-situ* spin transport phenomena.

## Results and Discussion

We present first a sample grown at an optimal coverage condition (500 Torr Ar/$H_2$ background at 935 °C for 2 hours). Verification of Pb intercalation is done using a standard assessment of the C 1s, Si 2p, and Pb 4f core level spectra in X-ray photoelectron spectroscopy (XPS) before and after intercalation.[22,65,67] We observe a characteristic reduction of peaks between 285 – 286 eV, a ~1.1 eV downshift in binding energy of features in the C 1s and Si 2p spectra associated with SiC, and emergence of asymmetric peaks at 136.5 eV and 141.1 eV related to metallic Pb, all of which indicate Pb intercalation within the EG/SiC gallery **(Figure S1)**. Furthermore, we can visualize the real-space sample surface through optical and electron microscopies taken in the same area **(Figure 1a-c, g)**. A region of darker optical contrast in the center of the image (region 1) also appears darker in backscattered electron (BSE) imaging and brighter in secondary electron (SE) imaging. Raman spectra from these two regions are distinct. Regions bright in optical and BSE images (region 2) show a series of broad spectral features between the Rayleigh line and 6H-SiC's folded transverse acoustic (FTA) mode at 150 $cm^{-1}$, which are present as both Stokes and anti-Stokes shifts **(Figure 1f,h)**.[70] These peaks are absent in region 1. The most identifiable peaks in region 2 are a peak at ~50 $cm^{-1}$ and a broader feature at ~90 $cm^{-1}$, which are similar to those reported previously for intercalated Pb on 6H-SiC.[70]

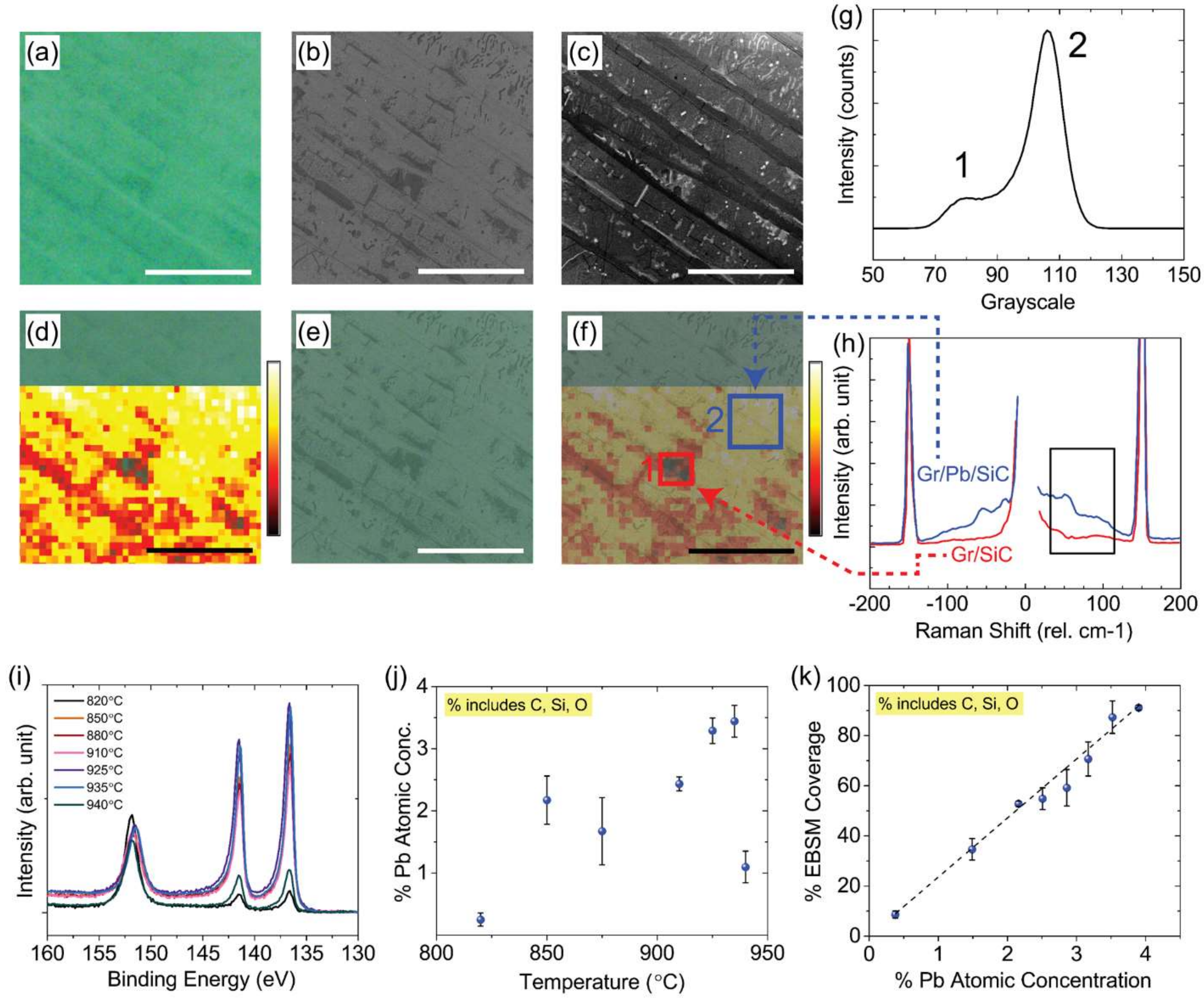

FIG 1: **Microscale microscopy and coverage analysis of Gr/Pb/SiC.** (a) Optical, (b) BSEM, and (c) SEM images within the same ~20×20 μm area. The same optical image is overlaid with (d) Raman spectroscopy mapping of the intensity in the ultra-low frequency range from 25-100 $cm^{-1}$, (e) the BSEM image (transparency 50%) and (f) the combined Raman spectroscopy mapping and BSEM, showing correlative features between all three. Scale bar is 10 μm in all images. (g) Grayscale histogram of (b) showing two distinct contrasts 1 and 2, which correspond to areas 1 (red box) and 2 (blue box) in (f). (h) The averaged ultra-low frequency spectra taken in the red (1, non-intercalated) and blue (2, intercalated) boxes showing a characteristic 2D-Pb signal for the blue box. Raman mapping is performed using the maximum intensity count in the black box. (i) High-resolution Pb 4f spectra overlaid for growth temperatures 820 °C to 940 °C. (j) Estimated atomic percentages of Pb from XPS versus synthesis temperature, indicating a maximum near 935 °C. (k) Relative atomic percentage of Pb (of entire composition including Si, C, O) from XPS vs. coverage percentages from EBSM, showing a tight linear correlation (r = 0.99938).

A systematic analysis under different synthesis conditions can provide further insight into the nature of this contrast and its relationship to the structure of Pb. Since the intercalation temperature affects surface arrangements and intercalation rate, and can induce deintercalation,[63] the Pb 4f spectrum and BSE images of a series of samples prepared under different annealing temperatures (820 °C – 940 °C) is presented in **Figure 1i**. Assuming the effective attenuation length[71] of the graphene overlayer is near equivalent for all photoionized Pb 4f electrons across all samples, the Pb 4f atomic concentration is a direct relative measure of the number of Pb atoms intercalated; this is plotted as a function of synthesis temperature in **Figure 1j**. Samples grown from 820 °C to 935 °C show a gradual increase in Pb concentration with increasing temperature and then a sharp decrease at 940 °C. This trend is unsurprising: it has already been reported for Pb and is common amongst various intercalant species, where a sharp decrease in the intercalant concentration marks the onset of deintercalation.[16,22,72] The thermal stability of intercalated Pb is of discussion at these elevated temperatures, as some reports see deintercalation as low as 700 °C.[63] We highlight that the CHet process, in contrast to other methods, provides a constant flux of sublimated Pb clusters (at a partial pressure of ~0.1 – 5 Torr in this temperature range) under an Ar overpressure which likely disfavors deintercalation.[73,74] Using this gradual increase, we can further plot Pb concentration against the percentage of region 2 contrast in BSE images taken on the same samples **(Figure 1k)**, obtaining a tight linear correlation. This suggests that the bright regions in **Figure 1b** are precisely where Pb has intercalated. Under this interpretation, optimal CHet synthesis conditions result in ~90% intercalant coverage, where intercalation is typically precluded at step edges which host increased graphene thicknesses[63,72] and a change in SiC crystallographic direction from (0001) to $(11\overline{2}n)$ along the edge that can inhibit diffusion.[75] Given that CHet-based Pb intercalates predominantly as a monolayer underneath mostly bilayer graphene, we calculate an areal density of $(1.0 \pm 0.09) \times 10^{15}$ $cm^{-2}$ Pb atoms from XPS in the optimal intercalation case, or a Pb/Si ratio of $(0.83 \pm 0.07)$.

We compare the atomic structure of CHet-based intercalated Pb in optimized samples against other Pb intercalation reports through LEED, STM, and STEM in a high angle annular dark field (HAADF) collection mode. To avoid surface contamination resulting from plasma-assisted intercalation, samples investigated by LEED and STM used a partial monolayer EG with ~10% exposed buffer regions. The samples otherwise have equivalent synthesis conditions, as described in the Methods. **Figure 2a** shows LEED diffraction captured at room temperature using a 96.4 eV incident electron beam. Beyond the known SiC and graphene 1$^{st}$ order diffraction spots, a set of superstructure spots arise around g(01), reflecting a quasi-(10×10) periodicity induced from Pb intercalation. A room-temperature STM image taken within a 50×50 $nm^2$ window is predominantly decorated with a periodic quasi-hexagonal modulation with a period of ~2.6 nm **(Figure 2b)**. Other observable features are bright spots which do not follow the hexagonal pattern and are likely protrusions from the underlying Pb layer (see **Figure S6**). A higher resolution image **(Figure 2c)** reveals a mottled triangular lattice pattern with the periodicity of bilayer graphene,[61] which suggests the longer length-scale quasi-hexagonal modulation originates from a lattice underneath the graphene. **Figure 2(d)** shows a fast-Fourier transformation (FFT) of the image showing two sets of sixfold patterns corresponding to lengths 0.246 nm and 2.64 nm (~ 10 * $a_{graphene}$) are observed. Similar observations of a 10×10 periodicity have been observed previously after Pb intercalation,[22,61,62,65] yet elucidation of the underlying surface reconstruction from LEED and STM alone has remained unclear. An initial model was that of a Moiré superperiodicity between the Pb and graphene lattices which corroborated LEED and STM micrographs[61,65] but diverges from both the rigid bias dependent contrast change in STM and photoemission studies.[67] Most recently, both a misfit dislocation network[67] and an amorphous phase[68] have been proposed in structural models to bridge discrepancies between characterization techniques. To further discern a structural model, we present atomic-scale, cross-sectional annular dark field scanning tunneling electron spectroscopy (ADF-STEM) (**Figure 2e**) and electron energy loss spectroscopy (EELS) (see **Figure S7**) on a Pb-intercalated sample along the (11-20) plane is SiC **(Figure 2e)**. A representative cross-sectional structure is shown beside the image. A hazy monolayer of Pb atoms is identifiable, weakly registered to SiC and sandwiched under trilayer graphene. A similar registry to SiC is seen with more clarity for CHet based Ga, In, and Sn intercalations which are understood to adopt a (1×1) structure on SiC(0001),[18] suggesting that similar local alignments may be present for Pb, with less columnar long-range registry.

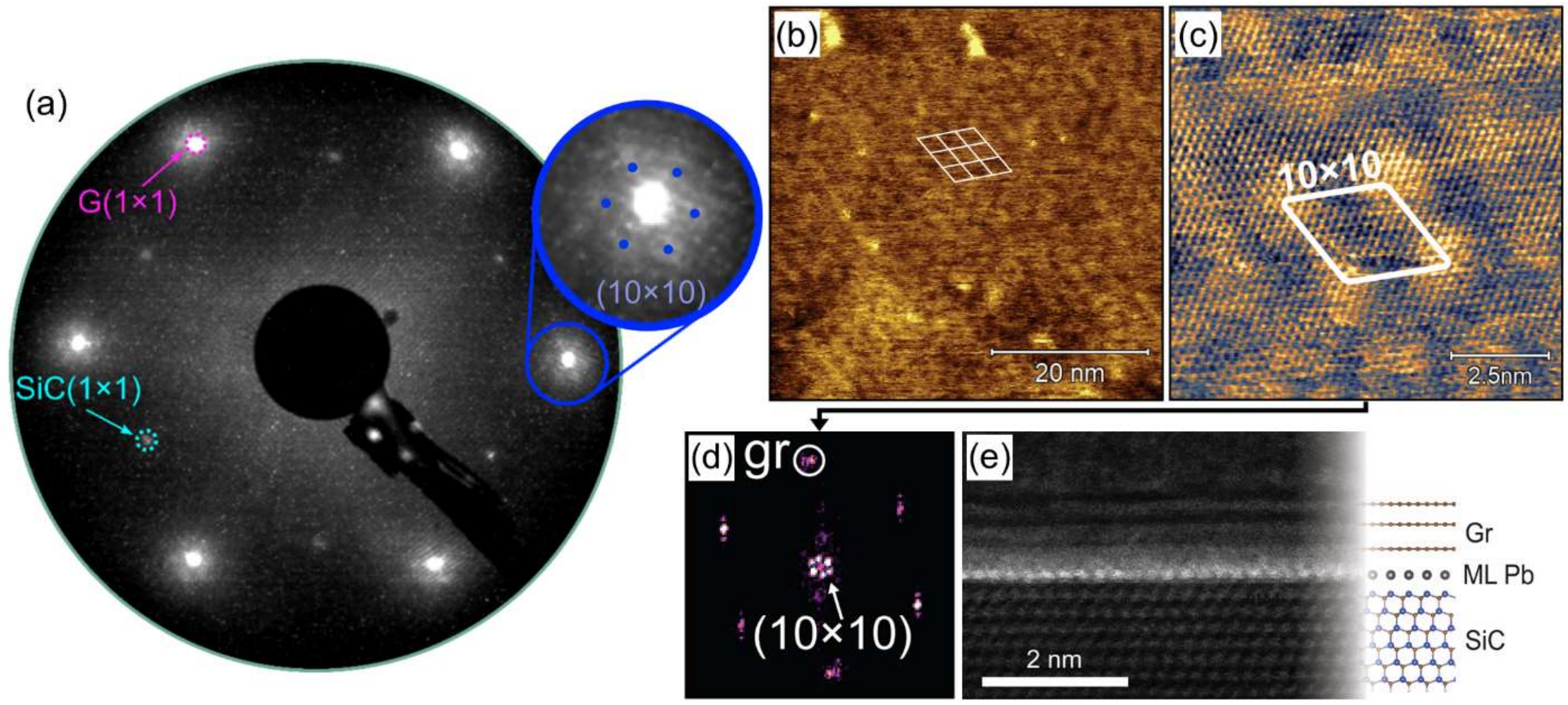

FIG 2: **Atomic structure of CHet-based Gr/Pb/SiC.** (a) Low energy electron diffraction (LEED) pattern taken at a electron kinetic energy of 96.4 eV. The image shows graphene spots, G(1×1), SiC(0001) spots, SiC(1×1) and diffraction spots from the Pb/Graphene superlattice which manifest as a (10×10) periodicity with respect to the graphene diffraction spots. (b) Surface morphology of Gr/Pb/SiC sample imaged by STM at (b) 50×50 nm$^2$, $V_{bias}$ = -1.05 V and $I_t$ = 0.85 nA and (c) 10×10 nm$^2$, $V_{bias}$ = -1.05 V and $I_t$ = 0.85 nA. (d) A 2D fast-Fourier transform (FFT) of the STM image shown in (c). (e) Cross-sectional STEM of a Gr/Pb/SiC sample. Pb atoms are seen sandwiched between graphene and SiC.

We performed angle-resolved photoemission spectroscopy (ARPES) to illuminate the electronic structure at the Pb-SiC interface. In the $K_{gra} \rightarrow \Gamma \rightarrow K_{Pb}$ cut **(Figure 3a)** characteristic π bands around the graphene K point ($K_{gra}$) appear, and also low-lying states around Γ from the valence band of SiC. Additional bands are also evident – one with upwards curvature from Γ to $K_{Pb}$ that splits 1.3 eV below the Fermi level, a band with upwards curvature from Γ to $M_{Pb}$ that disappears around ~0.8 eV below the Fermi level and an electron-like band with minima near −0.5 eV that splits around the $M_{Pb}$ point. These bands cross the Fermi energy and are visible on the experimentally measured Fermi surface **(Figure 3b)**. A similar band structure is seen in both Matta *et al.* and Schadlich *et. al.*, ,[65,67] however we do not observe the (10×10) replica cone bands seen in the latter report, likely due to a higher number of defects in plasma-treated graphene here, which could suppress the replica bands. We similarly observe that the additional bands are not present in SiC or graphene alone and also ascribe them to Pb, noting their likeness to a calculated DFT band structure **(Figure 3d,e)** with SOC for a (1×1) model[67]. We also note the close correspondence of the Fermi surface calculated for 1×1 coverage and the experimental ARPES Fermi surface **(Figure 3b,c)**; the Fermi energies used for the computed bands of Figures 3d,e closely match that for the Fermi surface of **Figure 3b**, within a few tens of meV. Two bands lose signal as they pass an avoided crossing **(Figure 3e)**; this drop in spectral weight is also observed in Matta et al.'s work.[65] To understand the origin of this truncation, we performed an ARPES simulation of monolayer Pb on top of SiC, the graphene layer being excluded in the model for computational simplicity **(Figure 3f)**. The gray lines in **Figure 3f** show the plain band structure, while the color-shaded regions show the simulated ARPES signal, which shows a close correspondence to experiment. By comparing to the orbital-projected band structure **(Figure S8)**, we see that ARPES predominately picks up initial states with $p_z$ character, while the "missing" bands carry mainly $p_x/p_y$ character. Through ARPES simulation, we deduce that both matrix elements and finite-resolution effects[76] play a role in the relative intensity of the different ARPES bands **(Figure S9)**.

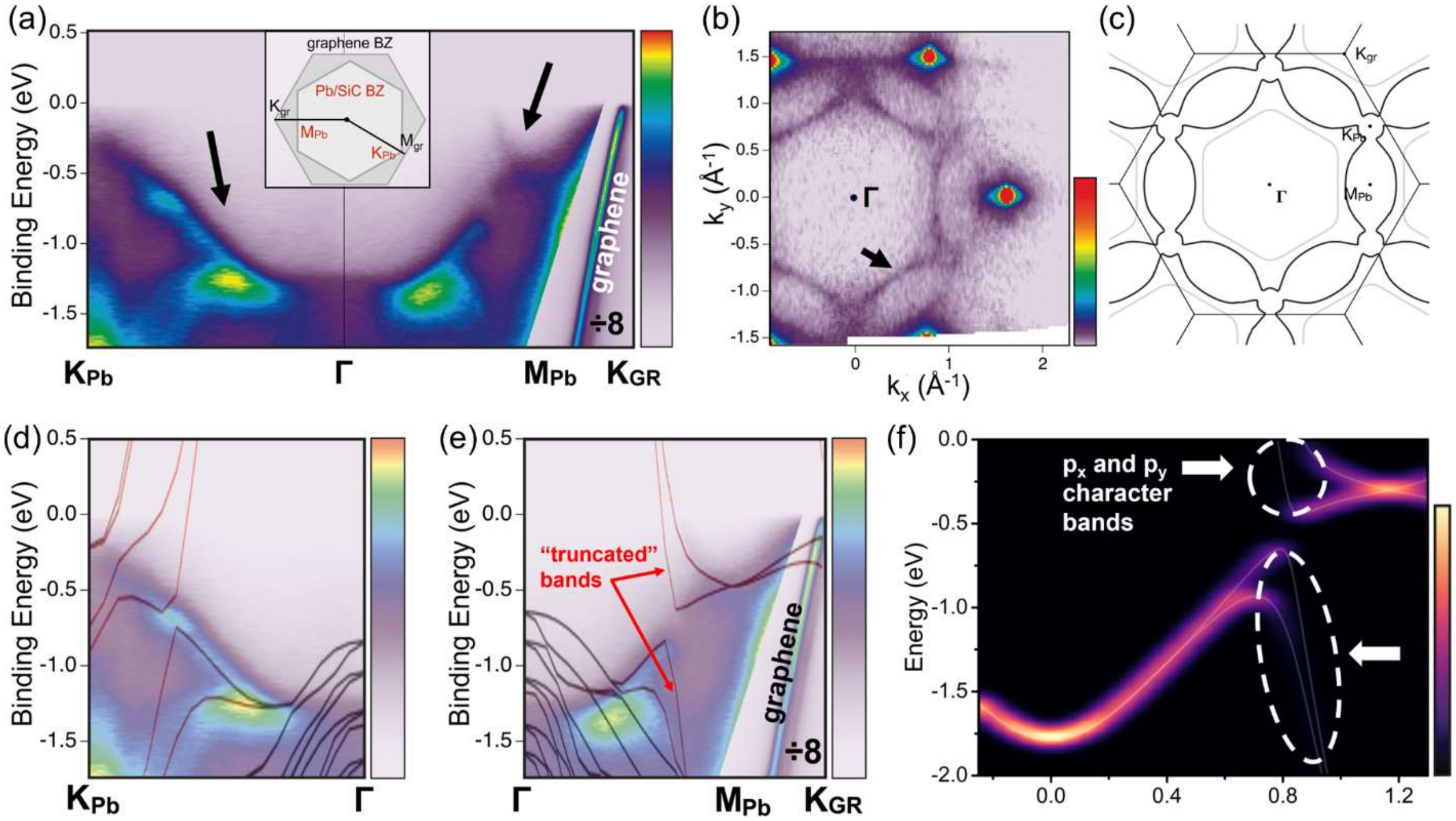

FIG 3: **Electronic band structure of Gr/Pb/SiC.** Measured ARPES (a) of Gr/Pb/SiC along a path crossing high symmetry points $K_{Pb} \longrightarrow \Gamma \longrightarrow M_{Pb} \longrightarrow K_{gr}$ in the Brillouin zone in the inset highlighted by dark solid lines. The color scale for signal intensity is provided for improved visualization. In addition, the intensity of the graphene band is scaled down by 1/8. Bands originating from Pb-SiC are identified by black arrows. The experimental (b) and calculated (c) Fermi surface of Gr/(1×1)Pb/SiC are well correlated, where axes in (b) denote the photoelectron's in-plane crystal momentum. The light gray contour in (c) corresponds to a band losing ARPES signal. (d) Measured ARPES map from (a) overlaid with calculated band structure of Pb-SiC for the $\Gamma - K_{Pb}$ cut and (e) $\Gamma - K_{Gra}$ cut, showing strong agreement apart from "truncated" bands between $\Gamma$ and $M_{Pb}$. (f) Simulated ARPES for monolayer Pb from $\Gamma$ to M. The k-path is extended slightly to ease comparison with experiment. Dashed white lines show where various bands lose their ARPES signal.

Considering the above characterizations, we can develop a structural model for CHet-based Pb intercalation that can reconcile the conflicting evidence for superlattice modulations and 1×1 registry. The hexagonal (111) plane of bulk Pb has a larger in-plane nearest neighbor distance (3.53 Å)[77,78] than the Si–Si distance within the (0001) plane of SiC (3.10 Å);[78,79] thus, a Pb monolayer lattice on the SiC(0001) surface registered 1×1 above Si will suffer compressive stress. To determine a thermodynamically favored stress relaxation mechanism which matches the observed STM results, we calculated the domain formation energy per unit area for a series of structures initialized with vacancy defects using density functional theory. Whereas this stress may be most efficiently relaxed within a supercell of two-dimensional hexagonal domains under a stress-free boundary condition, for computational efficiency and ease of interpretation, we calculate (without SOC) a series of one-dimensional supercells of parallel line defects that release Pb compressive stress unidirectionally **(Figure 4)**. Specifically, we examine $\sqrt{3} \times n\sqrt{3}$ supercells of the SiC primitive cell ($n$ ranging from 1 to 6) where the horizontal direction in the inset of **Figure 4a** aligns to a graphene lattice vector. To construct the relaxed domains, Pb atoms are initially registered 1×1 above the uppermost Si atoms and then $m$ rows ($m = 1,2,3$) of Pb atoms are removed to create a vacancy line defect that is then relaxed to release the in-plane stress. The domain formation energy per unit area is defined as:

$$\frac{E_{\text{domain}}}{\text{Area}} \propto \frac{E_{\text{removed}} + m\mu_{\text{Pb}} - E_{\text{no-removal}}}{n} \tag{1}$$

where $E_{\text{removed}}$ and $E_{\text{no-removal}}$ are the energies of the structurally relaxed supercells with and without the removal of Pb atoms, and $\mu_{Pb}$ is the Pb chemical potential referenced to bulk Pb. Since the chemical potential window for Pb to intercalate is narrow (Figure S13), any choice inside this window gives results similar to those of **Figure 4**. A negative formation energy signals that the corresponding structure is thermodynamically preferred to the 1×1 registry, reflecting the existence of compressive stress at 1×1 registry which can be relaxed through a network of vacancy line defects forming Frenkel-Kontorova domains.[80] Removing 2 rows of Pb is preferred to removing 1 or 3 whenever linear domains are thermodynamically favored because the double vacancy row can reform a near-triangular lattice after relaxation. The most favorable linear domain hosts a double vacancy row with n = 5 (preferred by ~0.001 eV/n over n = 6), suggesting the Pb(111) monolayer relaxes its compressive stress through a network of vacancy line defects forming Frenkel-Kontorova domains[80,81] with a domain width of 23.25 Å, rather than the smoothly evolving ideal Moiré structure of two weakly interacting incommensurate lattices. The observation of mottled, grain-like hexagonal domains ~2.64 nm in size is a good match to our calculations. Our result is in close agreement to Schädlich *et al.*[67] in their unrelaxed 2D model for monolayer Pb on SiC with a domain boundary network.

In addition, this domain-relaxation model does not conflict with the ARPES results aligning closely to the calculated band structure of the 1×1 system, nor with the cross-sectional TEM results showing apparent 1:1 Pb-Si registry. **Figure 4b** shows how far Pb atoms deviate from the ideal 1×1 Si-top position in the relaxed domain structure. If the underlying Pb atoms underwent a homogeneous stress relaxation with no influence from the Si substrate, then the variation in this deviation across the supercell (blue line) would be a straight line. Yet, Pb atoms evidently favor the Si-top position across the interior of the supercell, thus maintaining a near-1×1 registry across much of the domain, with Pb atoms at the domain boundary more rapidly shifting registry by means of a larger deviation change per site before re-assuming near-1×1 registry in the next domain. Both the low areal coverage of these transition regions and the large deviation change per site within them imply a low contribution to the ARPES signal from registries outside the preferred 1×1 Si-top position: the regions of large deviation change per site do not possess real-space periodicity on the scale of the reciprocal space sampled most effectively by ARPES, and ARPES may have difficulty picking up a small-wavevector superlattice band folding which may extend over only a handful of periods. Cross-sectional TEM provides highest contrast in regions where atomic columns align along the beam, i.e. the interiors of the relaxed domains, which have the lowest deviation change per site; for a hexagonal domain pattern such columns may occur across a significant fraction of the depth of the cross-section across the entire field of view.

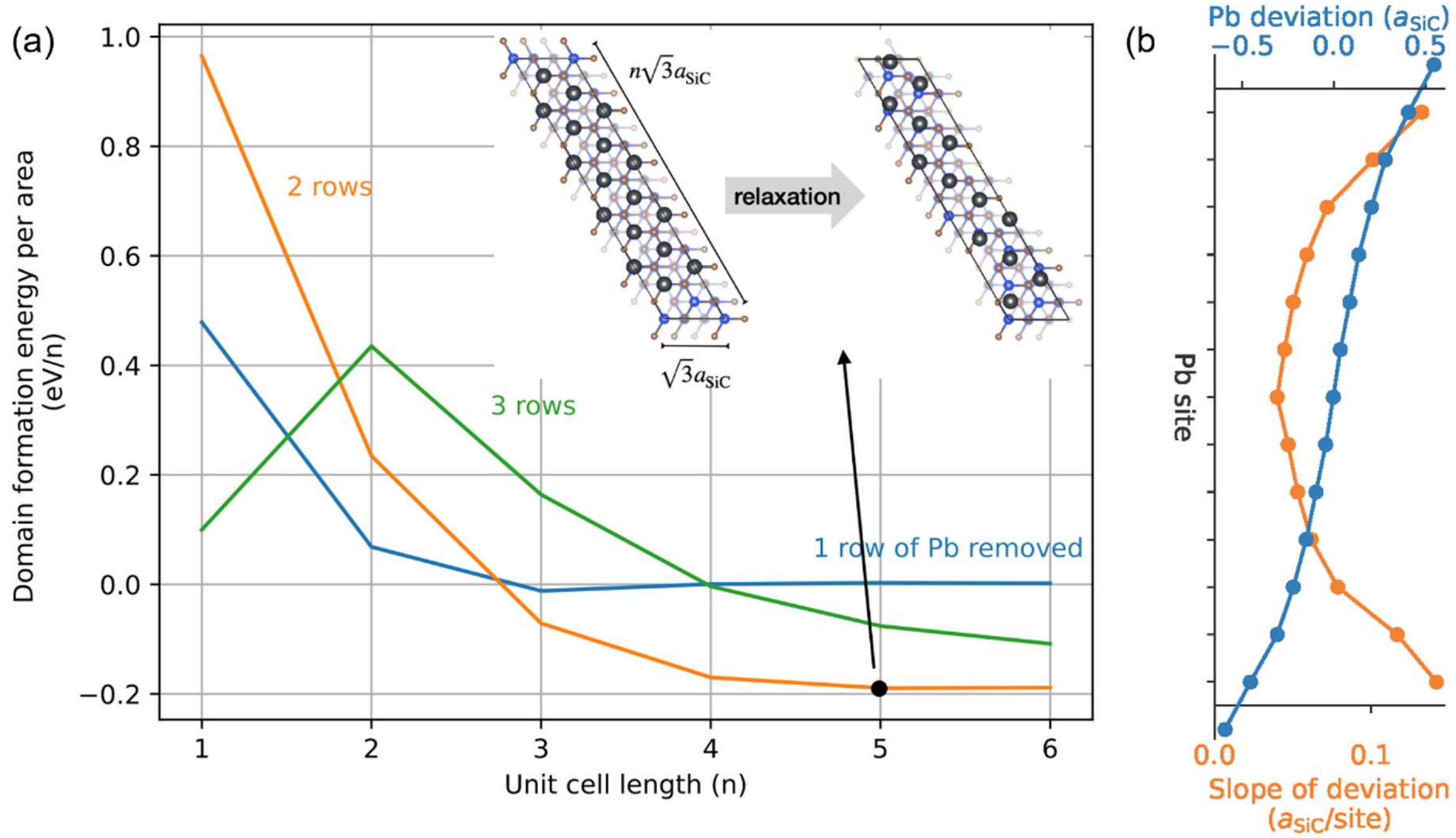

FIG 4: **Domain formation energy.** (a) Linear domains are created by relaxing striped supercells formed by removing 1, 2 or 3 rows of Pb atoms registered 1×1 above the Si sites of the uppermost SiC layer. The inset shows a √3×n√3 (n=5) SiC unit cell with 2 rows of Pb atoms removed; this case is energetically favored. Upon structural relaxation, the Pb atoms expand into the empty rows to relax the in-plane compression of the 1×1 registry but maintain near-Si-top registry in the interiors of domains. (b) The in-plane deviation away from the Si-top location for Pb atoms in the n=5 supercell. The preference for the Si-top registry is reflected in the flattening of the blue curve in the interior of the domain.

Room-temperature spin torque ferromagnetic resonance (ST-FMR) measurements provide insight into spin transport phenomena in CHet-grown EG/Pb/SiC. Using 50×10 μm soft ferromagnet permalloy/epitaxial graphene/Pb heterostructure devices (Py/EG/Pb), an external magnetic field (H) is applied to orient the magnetization of the ferromagnet, while a radiofrequency (RF) current generates a spin current in the EG/Pb layer, thereby producing spin accumulation at the graphene/Py interface **(Figure 5a, b)**. This creates a spin torque on the Py magnetization, causing it to precess and changing the resistance of the ferromagnet due to anisotropic magnetoresistance. Thus, the magnetization dynamics of Py can be measured as a rectified DC voltage ($V_{mix}$) produced by the mixing of the applied RF current and the varying resistance of the ferromagnet. The resonance phenomenon is characterized by analyzing the spectrum due to the mixing voltage while sweeping the external magnetic field (H) across the resonance at a fixed RF current **(Figure 5c)**. The spectrum is finally fit with a symmetric and antisymmetric Lorentzian contribution, as shown in **Figure 5d**. This results in a ratio of in-plane ($\tau_{||}$) and out-of-plane torque ($\tau_{\perp}$) (Eq. 1) that is proportional to the amplitude of the symmetric ($V_S$) and antisymmetric ($V_A$) Lorentzian contribution. This ratio is also related to the effective field in the in-plane ($H_{DL}$) and out-of-plane ($H_{FL}$) directions, since the torque is proportional to the magnetization ($\vec{m}$) and the effective field $\vec{H}_{eff}$ generated in the heterostructure ($\vec{\tau} = \vec{m} \times \overrightarrow{H_{eff}}$):[82]

$$\frac{H_{DL}}{H_{Oe}+H_{Fl}} = \frac{\tau_{||}}{\tau_{\perp}} = \sqrt{1 + \frac{4\pi M_{eff}}{H_{res}}} \frac{V_S}{V_A} \quad . \qquad (2)$$

Here, $H_{Oe}$ is the magnetic field generated by the flow of electrical current in the heterostructure, $H_{res}$ is the resonance field, and $4\pi M_{\mathrm{eff}}$ is the demagnetization field obtained from fitting $H_{res}$ with the Kittel ferromagnetic resonance equation. We note that the presence of the conducting EG layer in between the Pb and Py complicates the detailed analysis of ST-FMR data, in contrast to the typical ST-FMR heterostructure geometry where only a spin current generating layer and a ferromagnetic layer are involved.[82–87] The extraction of the in-plane damping-like spin torque efficiency, the out-of-plane field-like spin torque efficiency, and the spin Hall angle would require accurate modeling of the current distribution flowing through the parallel resistor network of Py, EG, and Pb. We can, however, obtain some insight into the charge-spin current conversion through the ratio $\frac{H_{DL}}{H_{Oe}+H_{FL}}$.To quantify the effect of the spin and charge currents in the ferromagnetic/EG/2D-Pb heterostructure, the ratio of effective fields in 2D-Pb is computed at different frequencies **(Figure 5e)** and compared its value with a H-intercalated graphene control sample. On average $\frac{H_{DL}}{H_{Oe}+H_{FL}} = 1.62 \pm 0.09$ in EG/Pb while it was $\frac{H_{DL}}{H_{Oe}+H_{FL}} = 0.97 \pm 0.19$ in the control sample. These values indicate 2D-Pb increases the effective field ratio, thus enhancing the charge to spin conversion in the heterostructure. The possible presence of both damping-like and field- like effective fields in 2D systems (such as the one produced by Rashba-like spin textures in 2D gases or in the surface states of a 3D topological insulator)[87–90] makes it difficult to precisely estimate the spin torque efficiency of 2D Pb. We expect in the future to perform a systematic study as a function of the ferromagnetic film thickness that will allow us to separate these contributions.

Finally, angle-dependent ST-FMR, where the angle (φ) between the current and the external magnetic field is changed, reveals the amplitude of the symmetric (S) and antisymmetric (A) components of the resonance peak as a function of φ (Figure 3e). Similar to other heavy metals in the 3D regime, the magnitude of the symmetric and antisymmetric components follows the usual $V_{mix} = V_{S(A)}\cos(\varphi)\sin(2\varphi)$ angle dependence.[87,88,91,92] This indicates that the spin polarization is completely in plane and perpendicular to the electrical current. This direction of spin polarization induced by electric current originates from the $C_{3v}$ symmetry of the 2D-Pb film. Considering the response equation $S_i = \chi_{ij}E_j$, where $S_i$ is the i-th component of spin density and $E_j$ is the electrical field in the j direction ($i = x, y, z$ and $j = x, y$), symmetry analysis based on Neumann's principle[93] suggests $\chi_{yx}$ as the only non-zero component (see Methods), which is indeed the component observed in experiments. This conclusion is further validated when comparing to a direct calculation of the current-induced spin polarization based on linear response theory for a realistic tight-binding model based on Wannier function from DFT calculations.[60] Yang et al.[60] provide a spin texture for 1×1 Pb/SiC at the Fermi surface, similar to the Rashba type of spin-split bands with concentric, reverse spin polarized circular bands. The corresponding values of $\chi_{yx}$ and $\chi_{zx}$ are shown in Figure S14 as a function of Fermi energy $E_F$. We find zero $\chi_{zx}$, as required by $C_{3v}$ symmetry, and non-zero $\chi_{yx}$, which shows strong dependence on the Fermi energy. The non-zero $\chi_{yx}$ provides an explanation of the observed spin-orbit torque in the ST-FMR measurements.

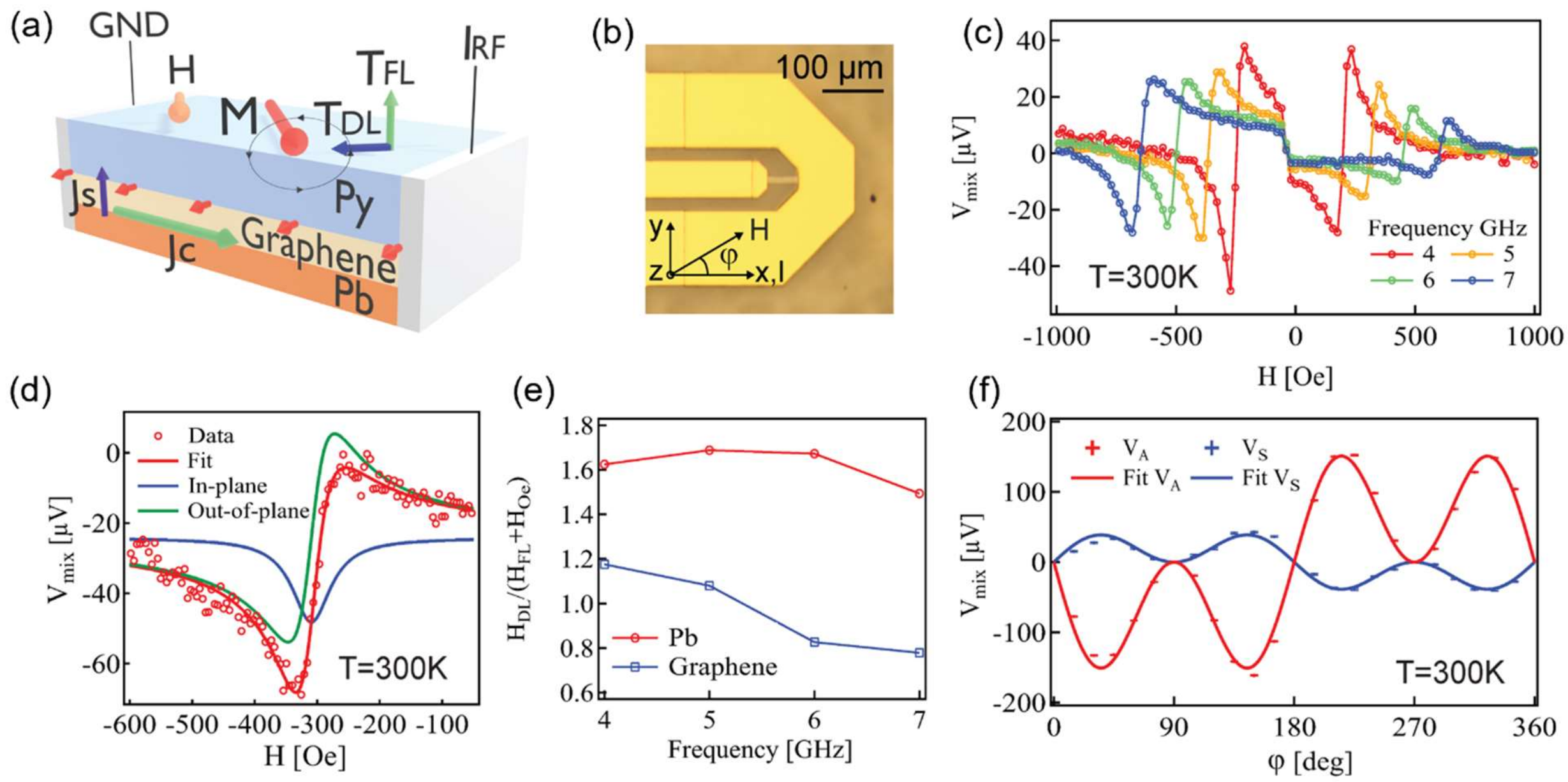

FIG 5: **Spin torque transport measurement of Gr/Pb/SiC.** (a) Schematic of the ST-FMR measurement. (b) Optical microscopy image of a 50×10 μm device used for ST-FMR experiments. We include the axis of our experiment and the directions of the current (I) and magnetic field (H). (c) ST-FMR spectra of a Py (6nm)/Gr/Pb heterostructure measured at room temperature using an RF signal of 20 decibel-milliwatts (dBm) ranging from 4 to 7 GHz. (d) ST-FMR spectra of the same heterostructure measured at 4 GHz showing the fit to the experimental data and the contribution of the in-plane and out-of-plane torques to the measured signal. (e) Normalized magnitude of the measured effective fields obtained from the fit of the data shown in (c) in graphene/Pb and its comparison with a graphene control sample without Pb. (f) Magnitude of the symmetric ($V_S$) and antisymmetric ($V_A$) components of the mixing voltage signal obtained by changing the angle (φ) between the current and the magnetic field. These values have been fit using $V_{mix} = V_{S(A)}\, cos(\varphi)\, sin(2\varphi)$ as mentioned in the main text.

## Conclusion

Through intercalation of Pb into EG/SiC via CHet, we can study Pb intercalation beyond the nanoscale with complimentary microscopy and spectroscopy techniques. Using these, we can achieve Pb coverages of up to 90% when synthesized at elevated temperatures above those of more conventional ultra-high vacuum setups. Akin to more recent detailed structural reports,[67] we find that CHet-based Pb exhibits a Frenkel-Kontorova domain boundary network due to lateral compressive stress relief within the (1×1) Pb/SiC model, which can be seen experimentally for lateral dimensions up to 200 $nm^2$ as a (10×10) superstructure pattern. Anticipating a large spin polarizability, we leverage uniformly intercalated Gr/Pb/SiC as a basal layer for a ST-FMR measurement, discovering a 1.5× increase in the effective field ratio over a hydrogenated sample and consistency with $C_{3v}$ symmetry. Hence, we conclude that Pb-intercalation via CHet serves as a viable route towards spintronic devices based on ultrathin heavy materials.

Multiple structural phases of CHet-based 2D metals are possible;[70] hence, alterations in the CHet setup (such as different pressures or cooling rates) may result in alternative phases of Pb such as the twisted honeycomb[66] or amorphous phases.[68] These phases may be topologically non-trivial,[94,95] which could significantly boost the spin-torque efficiency over what has been reported here. Other heavy elements, such as Bi[23,70], may also show non-trivial behavior,[96] representing a large charge-to-spin conversion. Future work could be focused on large-area growth of these phases and examining them below room temperature and/or with electrostatic gating in a similar setup as above. Future work could also focus on elucidating the effect of Frenkel-Kontorova domain boundaries on spin-torque measurements, which goes unexplored here.

## Methods

**Synthesis of Gr/Pb/SiC.** Atomically thin 2D-Pb is grown via confinement heteroepitaxy (CHet).[18] For this work, silicon carbide (SiC) (II-VI Inc.) is diced into 1 cm × 1 cm or 1 cm × 0.5 cm substrates and pre-cleaned by a 20-minute soak in Nano-Strip (VWR, 90% sulfuric acid, 5% peroxymonosulfuric acid, <1% hydrogen peroxide). Cleaned SiC wafers are then exposed to a standard etch process (1500°C, 700 Torr, 10% hydrogen bal. argon, 30 minutes). We subsequently sublimate silicon from the silicon carbide substrate and selectively grow, via anneal, nominally monolayer epitaxial graphene (EG) (1800 °C, 700 Torr argon, 10 min) or partially formed (~90%) monolayer EG (90% is performed with 1700 °C, 700 Torr argon for 40 minutes). We intentionally expose only nominally monolayer EG to a reactive plasma etch (500 mTorr, 150 sccm O2, 50 sccm He, 50 W, 1 minute) in a Tepla M4L plasma chamber to introduce defects into the graphene to ease the intercalation process[18]. Intercalation is achieved by heat treatment in a horizontal quartz tube (22 mm and 25 mm for inner and outer diameters, respectively) vacuum furnace, where lead powder (Sigma Aldrich, 99.999% trace metals basis, ~500 mg) is placed in an alumina crucible directly below a downward-facing EG/SiC substrate. Prior to heating, the tube furnace is evacuated and backfilled with ultra-high purity forming gas (96–97% Ar, 3–4% H2, to avoid surface particle accumulation, see **Figure S3**). Finally, the sample and Pb powder are heated to 800–940°C for 120 minutes under the forming gas environment at 500–700 Torr, with 200 sccm total gas flow. The sample is then cooled to room temperature within 30 minutes, using a fan. For comparison, we also examine hydrogenated EG, also known as quasi-free standing EG (QFEG), which is synthesized using previously established methods of hydrogen intercalation of EG on SiC.[72,97]

**X-ray Photoelectron Spectroscopy.** Samples are examined with a Physical Electronic Versa Probe II, using a monochromatic Al $K_\alpha$ X-ray source (hν = 1486.7 eV) at an incident angle of 45° from surface normal, with a radius of 100 μm and a concentric hemispheric analyzer. High resolution spectra are taken with a pass energy of 29.35 eV to 55 eV (Pb 4f and Si 2p) or 23.50 eV to 29.35 eV (C 1s) and acquisition times of ~540 seconds (C 1s), ~55 seconds (Si 2p), and ~337 seconds (Pb 4f). Fitting details are given in the supplemental for C 1s and Si 2p peaks. To quantify areal density, we use relative sensitivity factors based off Scofield photoelectric cross-sections and corrected for angular distribution and transmission function (based on our experimental setup).[98] For the effective attenuation length, we scale the intensity of our metallic Pb signal ($I_{Pb}$) using the equation:[99]

$$I_{Pb} = I_{Pb}^{\infty} exp\left\{-\frac{t}{[\lambda_e(E)\cos\theta]}\right\}$$

Where $\theta$ is the incident angle (45°), $t$ is the thickness of graphene, and $\lambda_e$ is the inelastic mean free path (IMFP) as a function of kinetic energy (E). The IMFP is determined from a previously reported empirical model using a modified Bethe equation:[100]

$$\lambda_e(E) = \frac{E}{{E_p}^2[\beta \ln(\gamma E) - \frac{C}{E} + \frac{D}{E^2}]}$$

$E_p$, $\beta$, $\gamma$, $C$, and $D$ are all parameters which we obtain from Amjadipour et. al.'s work.[100] We assume an areal density of $7.635 \times 10^{15}$ $cm^{-2}$ for bilayer graphene.

**Raman Spectroscopy and Microscopy.** A Horiba LabRam Raman system is used to perform spectroscopy with a laser wavelength of 532 nm at 4.1 mW focused through a 100x objective lens (NA 0.9). Double sweep spectra are taken with an accumulation time of 45 seconds and a grating of 600 grooves/mm. Raman imaging is done using the SWIFT ultra-fast imaging technique with a 1×1 μm pixel resolution. A flat baseline correction is applied to all data. For data in the supplemental, an additional correction is applied to subtract the SiC signal.

**Scanning Electron Microscopy.** Samples are imaged using a Verios 5 XHR SEM in immersion mode, equipped with TLD (secondary electron) and MD (backscattered electron) detectors. Images are taken at either 5000× or 10000× resolution, with a beam current of 0.4 – 3.2nA and voltage of 2.00keV at working distances between 2–5.6 mm. Histogram and threshold data are extracted using the ImageJ software.

**Low Energy Electron Diffraction.** Prior to LEED characterization, the Pb-intercalated sample was annealed at ~180 °C for 18 hours in UHV vacuum (~ $1\times10^{-10}$ mbar). After cooling to room temperature, LEED images were taken at a beam energy of 96.4 eV using a 22 mm rear-view LEED spectrometer (thoriated tungsten filament) and CMOS camera (4.92 megapixels).

**Scanning Tunnelling Microscopy and Scanning Tunnelling Spectroscopy.** For surface characterization at room temperature, the experimental sample was annealed at 250 °C for an hour under ultra-high vacuum (UHV) conditions (~$3\times10^{-10}$ mbar) to desorb potential surface contaminations. Subsequently, the sample was transferred to the analysis chamber (~$3\times10^{-10}$ mbar) for scanning tunneling microscopy (STM) characterization. STM analysis employed a Scienta Omicron VT-AFM, a UHV scanning probe microscope (SPM) designed for topographic and spectroscopic imaging of solid surfaces at sub-nanometer resolution. Gwyddion software was utilized for processing topographic images and performing 2-D FFT.[101]

**Scanning Transmission Electron Microscopy analysis.** Cross-sections from 2D-Pb samples were prepared by using a Helios G4 PFIB UXe DualBeam with a Xe+ plasma ion source. An electron beam at 5 keV and 6.4 nA was utilized to deposit a ~100 nm carbon protective coating. Then the Ga+ ion beam was used to deposit a 5 µm tungsten layer at 30 KeV. The samples were then prepared by performing a standard lift-out procedure and attached to a TEM half-grid. Finally, both sides of the samples were thinned in multiple steps by gradually lowering the ion beam voltage from 30 kV to 2 kV until the deposited tungsten is almost consumed, and the cross-section window appeared transparent in the electron beam image at 5 keV. The STEM images of the FIB cross-sections were performed in an FEI TITAN 80-300 KV HB Cubed Transmission Electron Microscope equipped with a double corrector for both image and probe. HAADF images were done at 200 kV with a dose rate of less than 50 e/Å$^2$ /sec using an in-column Fischione HAADF detector (model 3000). The beam convergence angle was set to 19.1 mrad with a 50 µm C2 aperture and the collection angles of 63-200 mrad at 91 mm camera length. The ADF STEM images were acquired at 300 kV and less than 50 pA screen current using Gatan ADF detector at 19.1 mrad with a 50 µm C2 aperture and the collection angles of 13-30 mrad at 91 mm camera length. The elemental mapping of the interface (in Figure S7) was performed using core-loss EELS at 300 keV, 29.5 mm camera length, and ~50 pA screen current using a direct electron detector Gatan K2 IS detector. EELS map was acquired at 0.0025 pixel time (0.0025 s per pixel) and 0.5 eV/channel electron dispersion, and 4 eV FWHM energy resolution. The EELS spectrum was denoised using Multi-Statistical Analysis for performing the elemental mapping. The Hartree- Slater cross-section method was applied to extract the core-loss signals, and power-law fitting was applied to remove the background from the EELS signal. STEM-EDX analyses were performed in a Thermo Fisher Scientific Talos F200X analytical microscope at 200 kV, equipped with an X-FEG source and four in-column super-x silicon drift detectors (SDD). STEM-EDX spectrum image (SI) datasets were collected with a dwell time of 50 µs with an image size of 1024×1024 pixels; 50 pA of beam current was used at spot size 10 nm to avoid damage to the Pb layer. The SI was collected, mapped, and analyzed using Velox software.

**Angle resolved photoelectron spectroscopy.** ARPES measurements were performed at the Microscopic and Electronic STRucture Observatory (MAESTRO) beamline at the Advanced Light Source at Lawrence Berkeley National Lab. The sample was annealed at 550 K for 1 hour before measurements to remove surface adsorbates. Measurements of 2D-Pb were performed using a photon energy of 110 eV. Photoemission spectra were collected by moving the sample around one angle while using the angle-resolved mode of a Scienta R4000 electron analyzer for the collection of the other angular axis.

**Theoretical and simulated band structure.** The spin-orbit coupled first-principle calculations are performed with VASP[102–104] at the PBE level[105,106] with a wavefunction energy cutoff of 600 eV and a Γ-centered 13×13×1 k-point mesh. The hopping terms of the tight-binding model of Pb p-orbitals and $p_z$ orbital of the top layer Si are extracted from the first-principle calculation with the code Wannier90.[107] This tight-binding model is then used by the code chinook[108] to simulate ARPES.

**Spin torque ferromagnetic resonance in 2D-Pb.** Six nanometers of permalloy ($Ni_{0.80}Fe_{0.20}$) were evaporated on top of CHet grown Pb. The films were later capped in situ with a 3 nm Al layer to prevent oxidation of the ferromagnetic layer, then subjected to standard lithography techniques, including etching with Ar and $SF_6$ to pattern the heterostructures into 50×10 μm devices which were then contacted using Ti/Au. The resistance of the devices was around 1 KΩ. The ST-FMR spectra was measured from 3 GHz to 10 GHz in a probe station equipped with a 40A GSG RF picoprobe, a GMW 5201 projected field electromagnet, a Keysight E8257D analog signal generator and a Keithley 2182A nanovoltmeter. The angle dependent measurements were performed at 4 GHz in a different setup using a rotating stage, a GMW 3470 electromagnet, an Anritsu MG3692C signal generator and a Keithley 2182A nanovoltmeter.

**Symmetry analysis of $\chi_{ij}$.** We apply symmetry operations $\hat{T}$ on the response equation $S_i = \chi_{ij} E_j$ where $S_i$ is the spin polarization and $E_j$ is the electrical field and the repeated indices means summation. Under the transformation, the spin polarization becomes $S_i' = \det(T)\ T_{ij} S_j$, the electrical field is $E_i' = T_{ij} E_j$ and $S_i' = \chi_{ij}' E_j'$ where $\det(T)$ is the determinant of $T_{ij}$. There is a determinant for $S_i$ because $S_i$ is a pseudovector. Writing the transformed response equation in terms of the original quantities gives $\det(T)\ T_{il} S_l = \chi_{ij}' T_{jk} E_k$, or equivalently, $S_i = \det(T)^{-1}\, T_{il}^{-1} \chi_{lj}' T_{jk} E_k$. Thus, the transformed response coefficient is determined by $\chi_{ik} = \det(T)^{-1}\, T_{il}^{-1} \chi_{lj}' T_{jk}$. For any symmetry of the system, we require $\chi_{ij}' = \chi_{ij}$ and the response coefficients need to satisfy:

$$\chi_{ik} = \det(T)^{-1}\, T_{il}^{-1} \chi_{lj} T_{jk}.$$

The current system has the $C_{3v}$ symmetry. The three-fold rotation $C_3$ about the z-axis can be given by the matrix:

$$C_3 = \begin{pmatrix} \cos 2\pi/3 & -\sin 2\pi/3 & 0 \\ \sin 2\pi/3 & \cos 2\pi/3 & 0 \\ 0 & 0 & 1 \end{pmatrix}.$$

It gives $\chi_{zx}, \chi_{zy}, \chi_{xz}, \chi_{yz}$ to be zero, $\chi_{xx} = \chi_{yy}$ and $\chi_{xy} = -\chi_{yx}$. For the mirror symmetry, we choose the zx plane as the mirror plane, and the mirror symmetry operation $m_y$ is:

$$m_y = \begin{pmatrix} 1 & 0 & 0 \\ 0 & -1 & 0 \\ 0 & 0 & 1 \end{pmatrix}.$$

Only the responses coefficients $\chi_{ij}$ with one of the indices as y are nonzero, namely, only $\chi_{xy}, \chi_{yx}, \chi_{yz}, \chi_{zy}$ are nonzero. Combing the requirements from these two symmetries, we have nonzero coefficients $\chi_{yx} = -\chi_{xy}$ for the current-induced spin polarization.

**Numerical method for the calculation of current-induced spin polarization.** We apply linear response theory[109] to calculate $\chi_{ij}$, which is given by:

$$\chi_{ij} = -\frac{1}{2\pi}\int \frac{d^2\vec{k}}{(2\pi)^2}\mathrm{Tr}s_i G^R(\vec{k},E_f)\Gamma_j(\vec{k},E_f)G^A(\vec{k},E_f).$$

The integral of momenta is over the Brillouin zone. $s_i$ is the spin operator. $G^R(G^A)$ is the retarded (advanced) Green's function and given by:

$$G^{R(A)}(\vec{k},E) = \left(E - H(\vec{k}) - \Sigma^{R(A)}(E)\right)^{-1}.$$

The $H(\vec{k})$ is the tight binding Hamiltonian for 2D Pb on SiC and is constructed from the Wannier interpolation of the density functional theory with three p-orbitals of Pb atoms and one $p_z$ orbital of SiC[60]. $\Sigma^R(E)$ is the self-energies from the short range disorder and self-consistently given by:

$$\Sigma^R(E) = n_i V_0^2 \int \frac{d^2\vec{k'}}{(2\pi)^2} G^R(\vec{k}',E).$$

$n_i$ is the disorder density and $V_0$ is the disorder strength. We take $n_i V_0^2 = 0.05eV^2\text{Å}^{-2}$ for the calculation. $\Sigma^A(E) = \left(\Sigma^R(E)\right)^{\dagger}$. $\Gamma_j(\vec{k},E)$ is the vertex correction taken in the ladder approximation and self-consistently given by:

$$\Gamma_j(\vec{k},E) = J_j(\vec{k}) + \gamma_j(E)$$

$$\gamma_j(E) = n_i V_0^2 \int \frac{d^2\vec{k'}}{(2\pi)^2} G^R(\vec{k}',E)\left(J_j(\vec{k'}) + \gamma_j(E)\right)G^A(\vec{k}',E).$$

where $J_i(\vec{k}) = -e\partial_{k_i}H(\vec{k})$ is the current operator and $-e$ is the electron charge.

**Conflicts of Interest**

The authors declare no competing financial interest.

**Acknowledgements**

This work is supported by NSF award DMR2002651 and the Penn State MRSEC Center for Nanoscale Science via NSF award DMR2011839. Raman, SEM, and XPS measurements were performed in the Materials Characterization Laboratory in the Materials Research Institute at Penn State University. LEED measurements were performed at the Laboratory of Physical Sciences at the University of Maryland. STM was performed at the University of Texas at Dallas supported by NSF award DMR2002741. Electron microscopy was performed at the Canadian Centre for Electron Microscopy, a Canada Foundation for Innovation Major Science Initiatives funded facility (also supported by NSERC and other government agencies). EM work was funded by the US AFOSR Award FA9550-19-1-0239 and the NSERC (Natural Sciences and Engineering Research Council of Canada) Discovery Grant program. This research used resources of the Advanced Light Source, which is a DOE Office of Science User Facility under contract no. DE-AC02-05CH11231. A portion of this research was conducted at the Center for Nanophase Materials Sciences, which is a DOE Office of Science User Facility. A. Vera is supported by the Alfred P. Sloan Foundation G-2019-11435. T. Bowen is supported by an NSF Graduate Research Fellowship (ID -like). B. Zheng and Y. Wang are supported by 2DCC-MIP under NSF cooperative agreement DMR-1539916 and DMR-2039351. S.Y. Kim was supported by NSF award DMR2002741. We thank Y. Ou and J. Zhu for helpful discussions, G. Zheng, A. Baddorf, and A.P. Li for providing access and help with LEED, C. Whittier for help taking the FIB cross-sections and A. Sengupta for providing access to the apparatus used in ST-FMR measurements. J.C.K and A.L.F acknowledge R. E. Butera for tool access.

**Land Acknowledgement:** The Pennsylvania State University campuses are located on the original homelands of the Erie, Haudenosaunee (Seneca, Cayuga, Onondaga, Oneida, Mohawk, and Tuscarora), Lenape (Delaware Nation, Delaware Tribe, Stockbridge-Munsee), Monongahela, Shawnee (Absentee, Eastern, and Oklahoma), Susquehannock, and Wahzhazhe (Osage) Nations. As a land grant institution, we acknowledge and honor the traditional caretakers of these lands and strive to understand and model their responsible stewardship. We also acknowledge the longer history of these lands and our place in that history. (https://equity.psu.edu/equity-at-penn-state/penn-state-resources/acknowledgement-of-land)

## Supporting Information

### Additional Structural Data for Gr/Pb/SiC

The structural evolution of a sample in CHet can be monitored via changes in the C 1s, Si 2p, and Pb 4f core levels high resolution as measured by X-ray photoelectron spectroscopy (XPS) (Figure S3) and Raman spectroscopy (Fig. S4) via modifications in the D/G ratio. The C 1s core level spectra for monolayer EG exhibits buffer layer peaks (S1 and S2, 285.1 eV and 285.6 eV, respectively), $sp^2$ or graphene (284.5 eV) and the bulk SiC substrate (283.5 eV). Plasma treatment introduces oxygen-based defects into EG, resulting in additional peaks related to C-OH (286.8 eV) and C=O (288.4 eV) bonding.[1,2] After intercalation, peaks assigned to the buffer layer become undetectable, the graphene peak increases in intensity, and the peak associated with SiC shifts to lower binding energy (282.4 eV) due to band bending.[3] These features all indicate the interface is passivated[4] and a second layer of graphene has formed, verified in transmission electron microscopy (TEM) post CHet. The passivation of the Si face of SiC, along with the formation of bilayer graphene, are consistent with our previous report on 2D-Ga, 2D-In, and 2D-Sn.[5]

Raman spectroscopy (Figure S4) details the evolution of graphene throughout CHet[6]. At each step we see peaks which we identify as the D (~1360 $cm^{-1}$), G (~1600 $cm^{-1}$) and 2D (~2720 $cm^{-1}$) peaks for graphene.[6] The D/G ratio average (near zero), G/2D ratio average (1.27), and FWHM average of the 2D peak (33 $cm^{-1}$) for pristine EG (EG before plasma treatment) suggest a high-quality monolayer of EG has formed on SiC. The D/G ratio increases substantially after plasma treatment, indicating a significant increase in defect density in the graphene lattice. The decrease in the D/G ratio after intercalation implies healing of the graphene layer, and the average G/2D ratio (1.6) and 2D FWHM (70 $cm^{-1}$) hint that the graphene is in the few-layer regime. Notably, the D/G ratios for Pb-intercalated EG are larger than for other CHet-based 2D metals[5] which can contribute to a larger 2D FWHM, but the absence of extra peaks associated with C-O bonding in XPS suggests these defects are minimal.

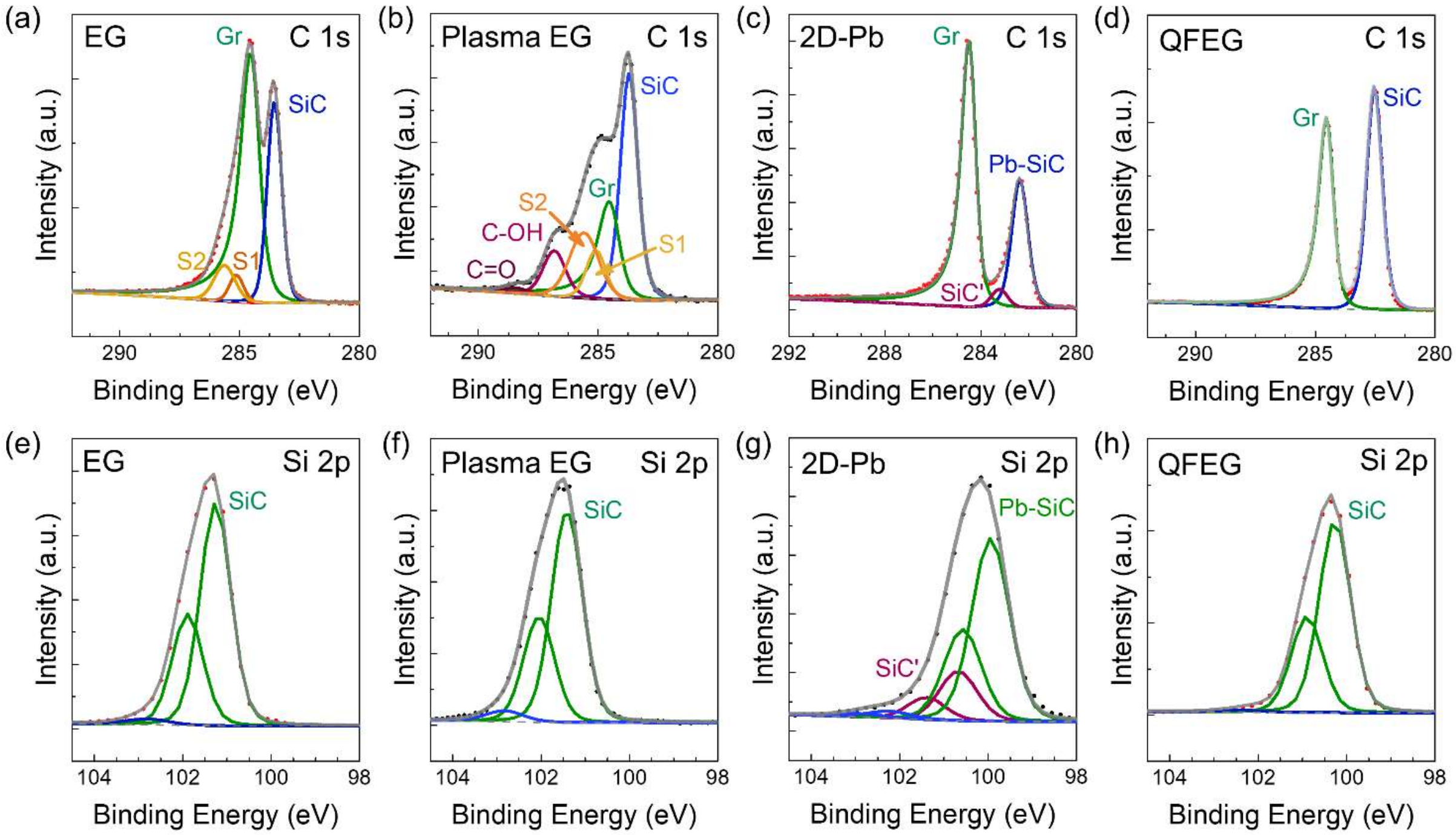

FIG S1: **XPS for CHet of Gr/Pb/SiC.** High resolution C 1s spectra from XPS for (a) EG, (b) plasma treated EG, (c) 2D-Pb, and (d) quasi-free standing EG. (e-h) Similarly, high resolution Si 2p spectra for samples mentioned previously. For C 1s spectra, a pass energy of 23.5 to 29.4 eV is used, with acquisition times between 3 and 9 minutes, and an energy step of 0.1 or 0.125 eV. For Si 2p spectra, a pass energy of 29.4 to 55 eV is used, with acquisition times between 45 and 90 seconds and an energy step of 0.125 or 0.2 eV. All spectra are charge corrected (sp2 at 284.5 eV) and fit with Voigt lineshapes defined in the CasaXPS software.[7,8]

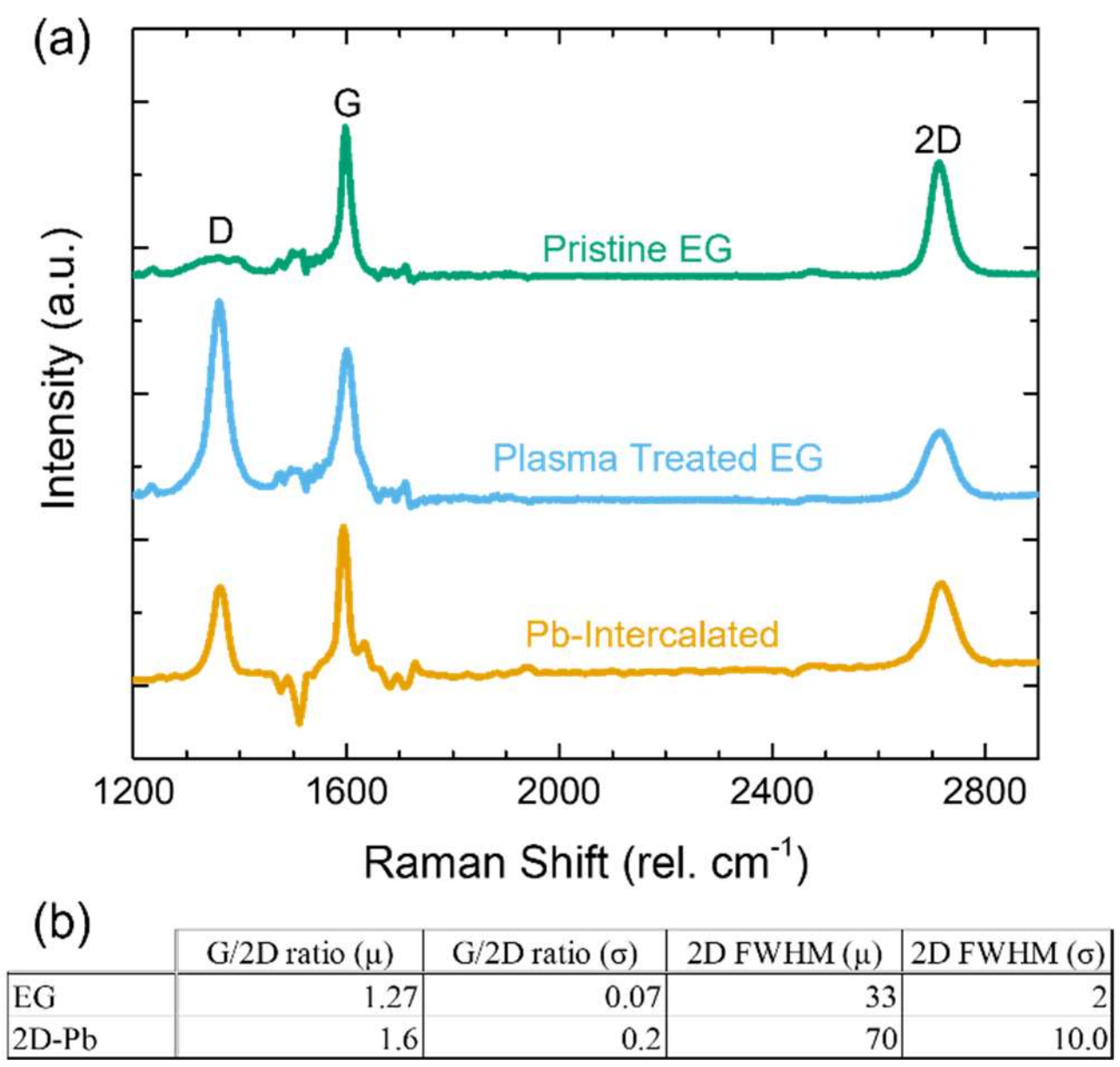

| | G/2D ratio (μ) | G/2D ratio (σ) | 2D FWHM (μ) | 2D FWHM (σ) |
|---|---|---|---|---|
| EG | 1.27 | 0.07 | 33 | 2 |
| 2D-Pb | 1.6 | 0.2 | 70 | 10.0 |

FIG S2: **Raman for CHet of Gr/Pb/SiC.** Raman spectroscopy (a) averaged over a pristine EG sample, a plasma treated EG sample, and a Pb-intercalated sample, with the SiC background signal subtracted. The D/G peak ratio reduces upon Pb intercalation after plasma treatment; however, the graphene remains more defective than for other metals. (b) Average (μ) and standard deviation (σ) for the G/2D ratio and 2D FWHM for EG and Pb-intercalated (aka 2D-Pb) samples. Despite being defective graphene, the 2D FWHM for 2D-Pb and G/2D ratio remains within the few-layer regime.

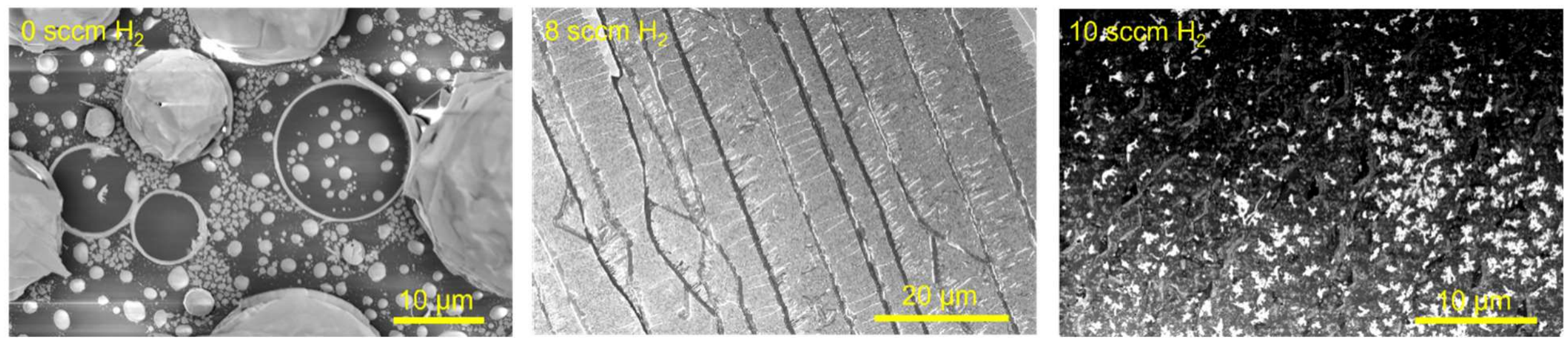

FIG S3: **Hydrogen inclusion for 2D-Pb synthesis.** SEM images taken for 2D-Pb samples grown with 0 sccm H2 flow, 8 sccm H2 flow, and 10 sccm H2 flow. Large particle formation can be seen for the case of no H2 inclusion, and white “flakes” can be seen for the case of 10 sccm H2 inclusion.

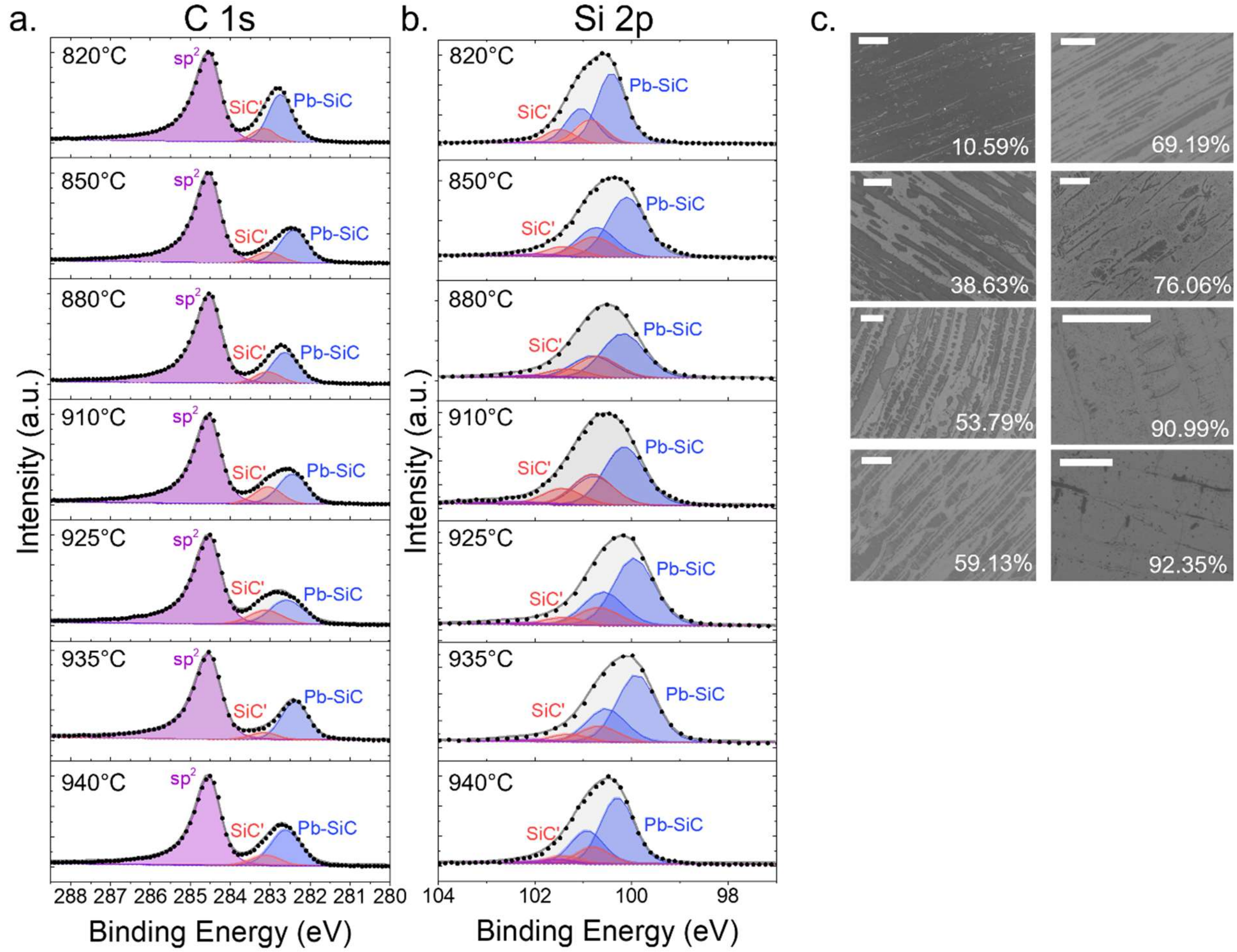

FIG S4: **Additional XPS and BSEM images of EG/Pb/SiC.** (a) High-resolution C 1s spectra; (b) Si 2p spectra from XPS for synthesis at 820°C to 940°C; (c) Electron backscatter microscopy (EBSM) images for coverage percentages from 10% to 93% (scale bar is 10 μm in all images). Expectedly, the largest coverage percentages yield the largest redshift of the SiC peak to lower binding energy. We find similar peaks as in Schadlich et. al.[9] (Bulk$_2$ and Bulk$_{10\times10}$ in their report, corresponding to SiC' and Pb-SiC here), further suggesting a similar structure is present here.

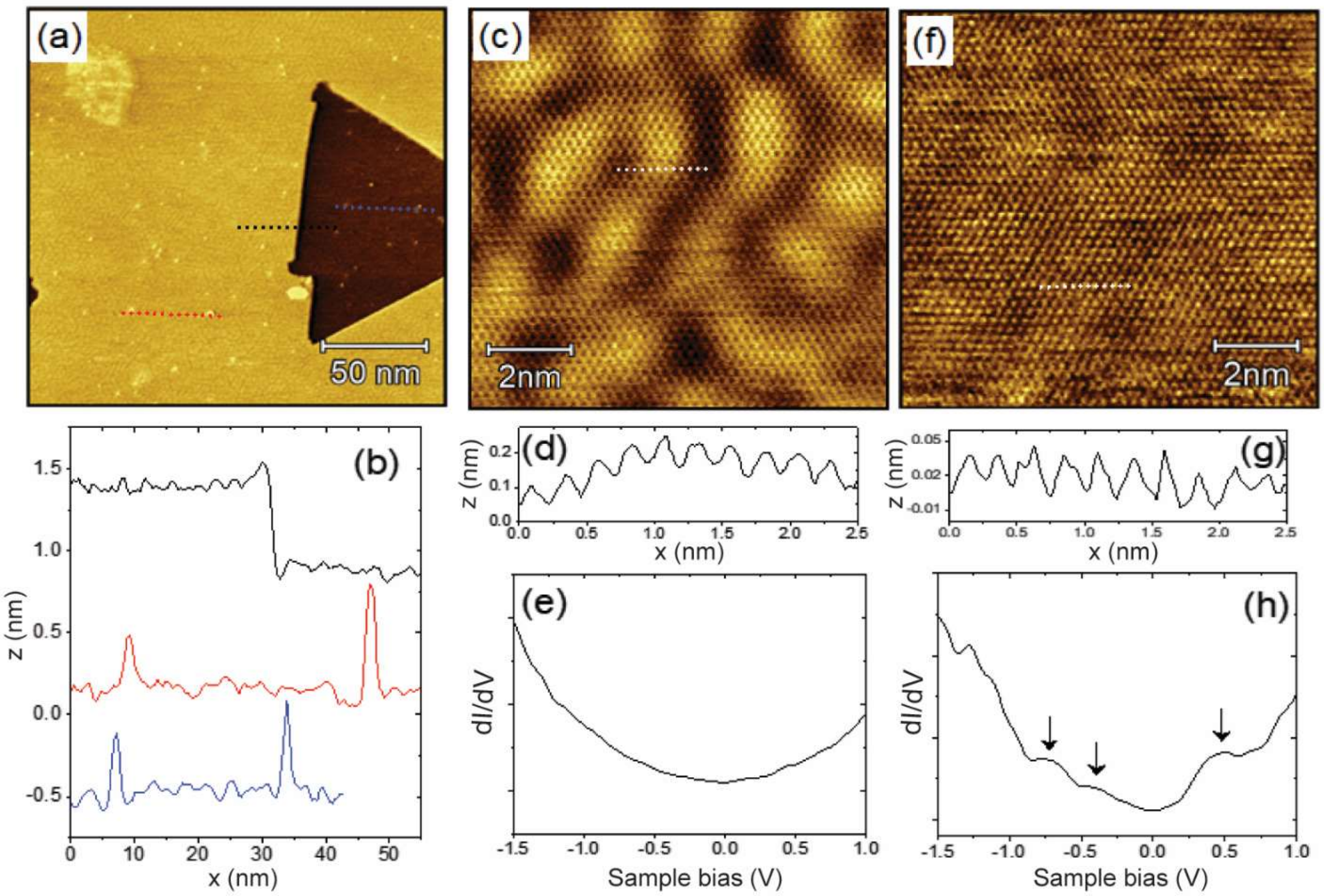

FIG S5: **Room temperature atomic resolution STM images and STS results of EG/Pb/SiC.** (a) Surface morphology of Gr/Pb/SiC sample in a 200 × 200 $nm^2$ window ($V_{bias}$ = 1.75 V and $I_t$ = 0.65 nA), where the surface is predominantly mottled with the (10 × 10) superlattice seen in Figure 2 but is also decorated with protrusions (see Fig S4) and shows a triangular shaped depression, related to SiC layers underneath. (b) Height profiles corresponding to dashed lines in (a), showing protrusions range from ~3 Å to 6 Å, and the triangular depression is ~7 Å (about 3 tetrahedral layers of SiC). (c, f) Atomic resolution images of two different locations on the same sample in a 10 × 10 $nm^2$ window showing irregular and weak patterning, respectively, similar to Hu et. al.'s report.[10] (d, g) Associated height profiles along dashed white lines in each scan, and (e, h) averaged STS spectra taken at each location. The density of states shows a minimum at 0 V. Black arrows point to states seen in these regions.

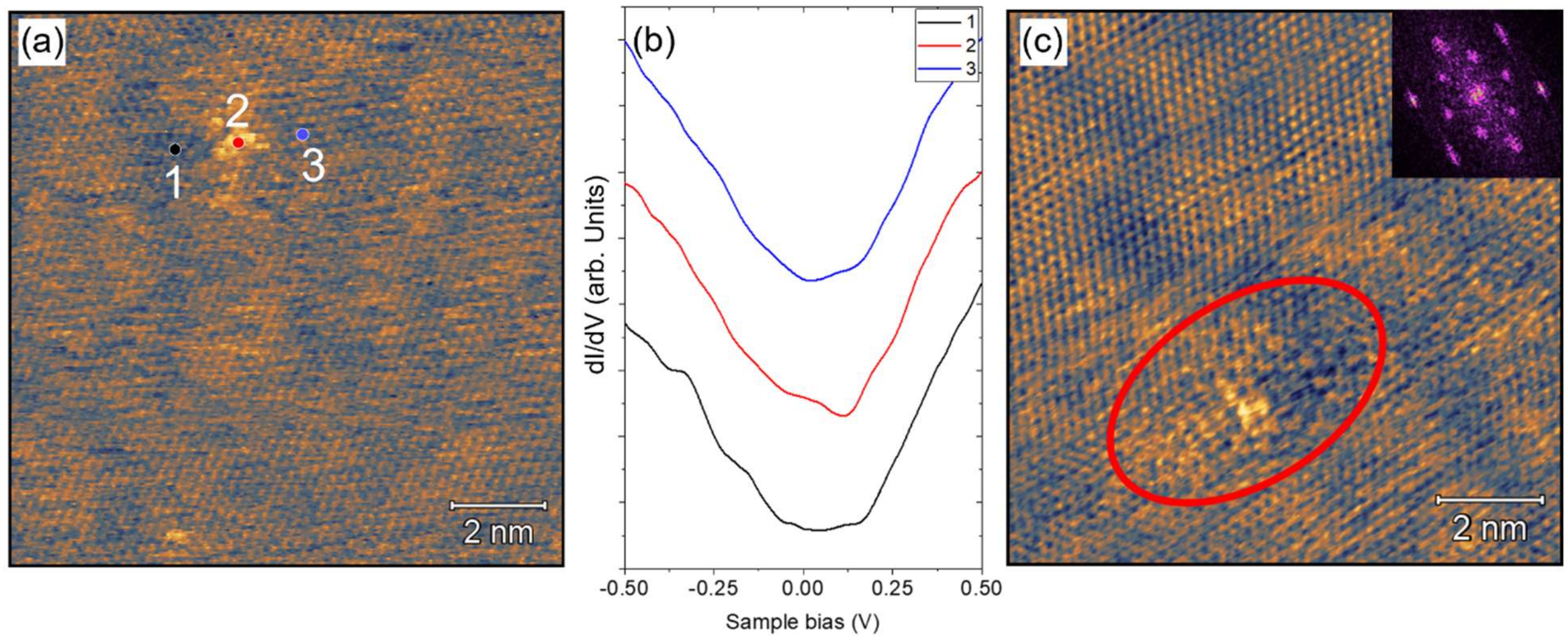

FIG S6: **Protrusions within EG/Pb/SiC.** (a) A 10 × 10 $nm^2$ window with a single protrusion, as seen in FIG S5, centered around point 2. Note that the graphene lattice is still continuous over the protrusion, suggesting an origin beneath the topmost graphene sheet. (b) STS spectra taken at points 1-3 in (a). The density of state minimum for point 2 shifts to ~0.1 V. Scans are offset vertically for visualization. (c) Another 10 × 10 $nm^2$ window on the same sample with another protrusion, where a change in the superstructure pattern is more evident. Inset is the FFT of the scan near the protrusion, showing a different hexagonal pattern from Figure 2(d) with length-scale 0.45 nm ($\sim\sqrt{3}\ a_{graphene}$).

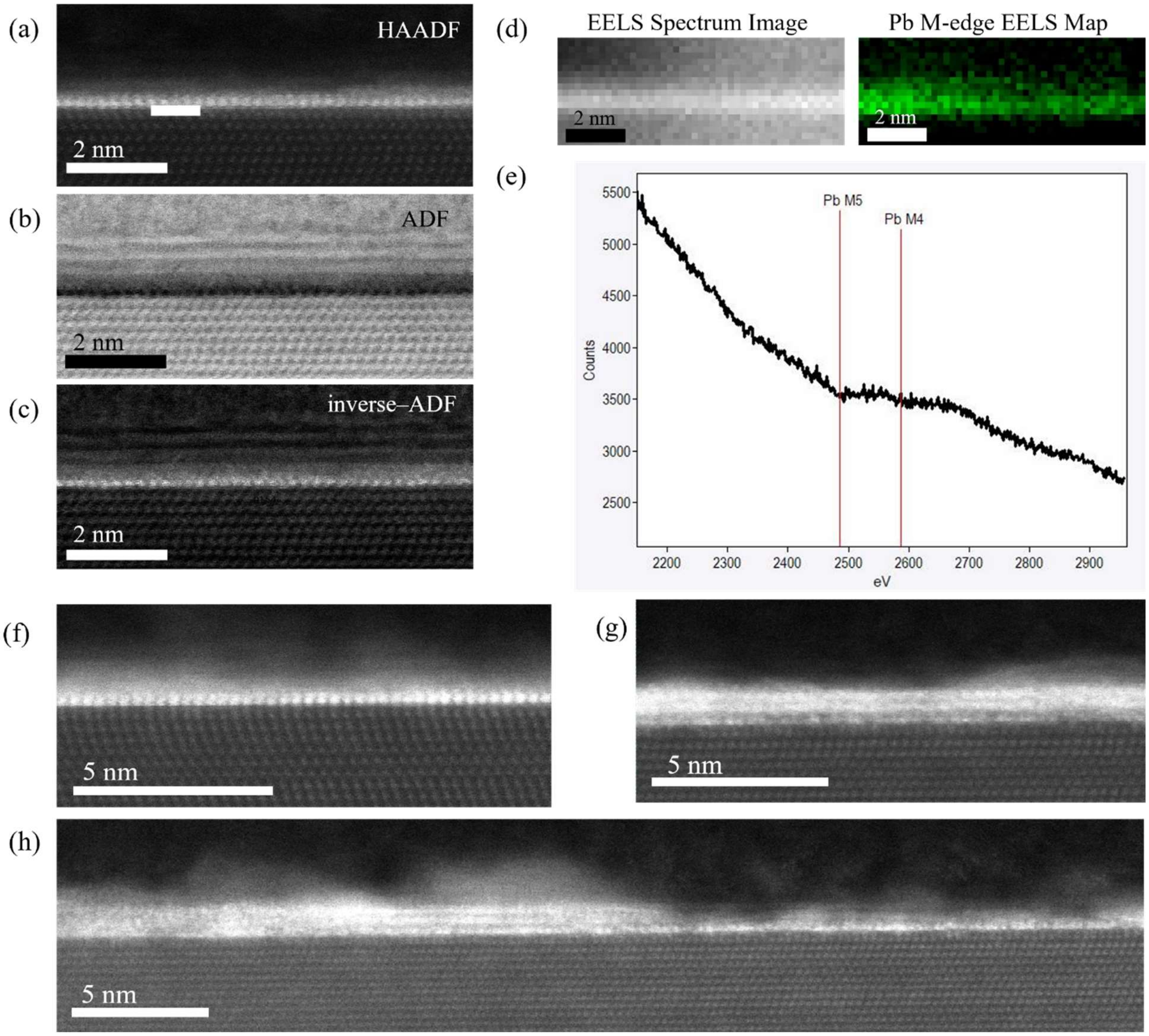

FIG S7: **Additional TEM of EG/Pb/SiC.** Cross-sectional scanning transmission electron microscopy (STEM) of 2D-Pb using (a) high-angle annular dark field (HAADF), (b) annular dark field (ADF), and (c) inverse annular dark field imaging. (d) Electron energy loss spectra (EELS) mapping for the full spectrum and for the Pb-M edge; (e) EELS spectra at a point with Pb inclusion. Cross-sectional STEM-HAADF images of (f) one, (g) three, and (h) one to three layers of Pb on SiC. These 3-layer regions may correspond to the protrusions seen in STM. Plumes of white in these images may be due to beam damage or instability during imaging.

**Band structures for 1×1 registry.**

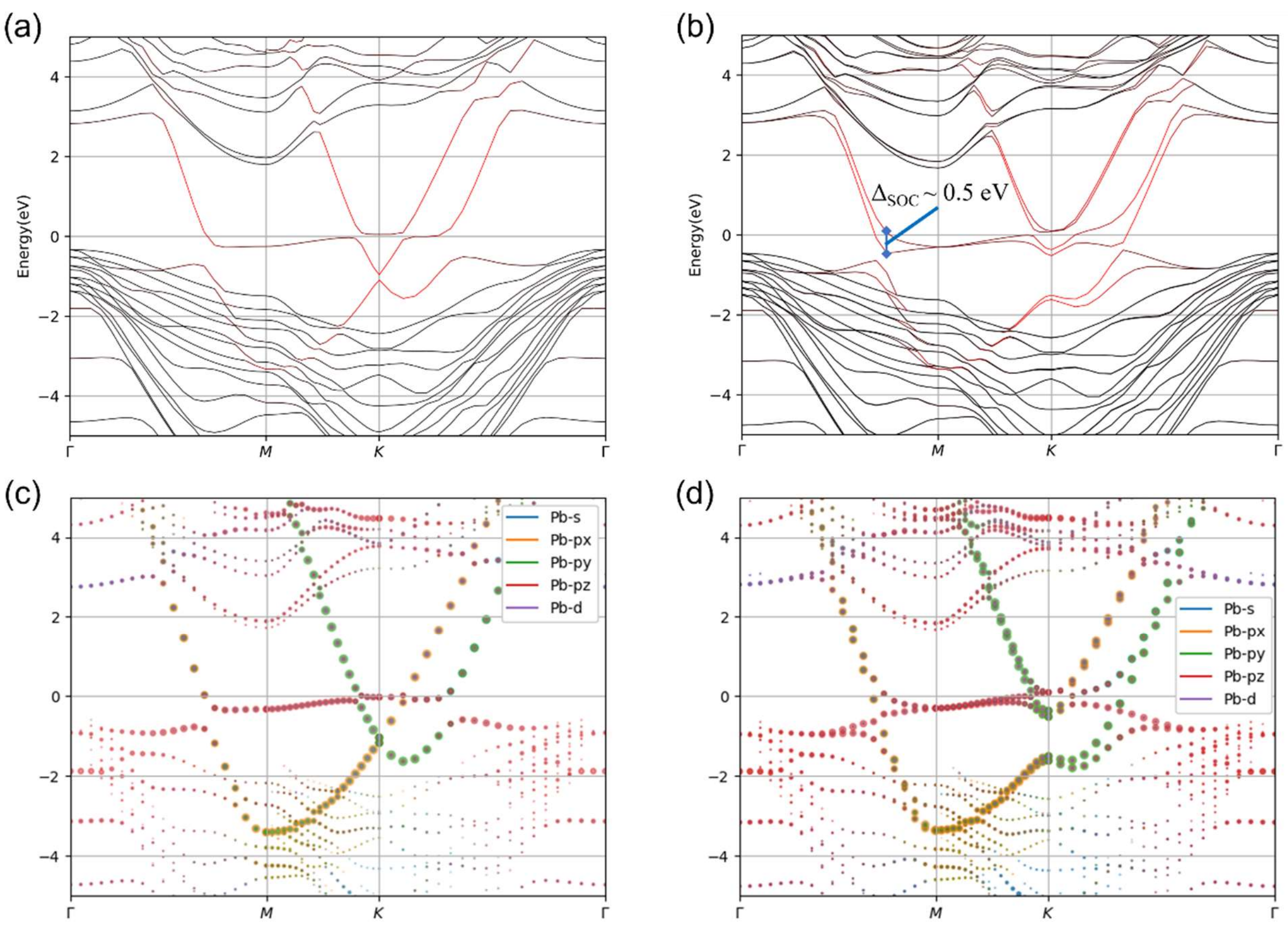

FIG S8*:* **Calculated band structures.** (a,c) Band structures without spin-orbit coupling. (b,d) Band structures with spin-orbit coupling. The Fermi level is at 0 eV. The red color in (a,b) is proportional to the amplitude of the projection to all Pb orbitals. The size of dots in (c,d) is proportional to the amplitude of the projection to the corresponding Pb-s, $p_x$, $p_y$, and $p_z$ orbitals. A steep band between the Γ and M points which has mainly $p_x$ and $p_y$ character may not be visible in experimental ARPES.

**Explanation of the band "truncation" in ARPES.**

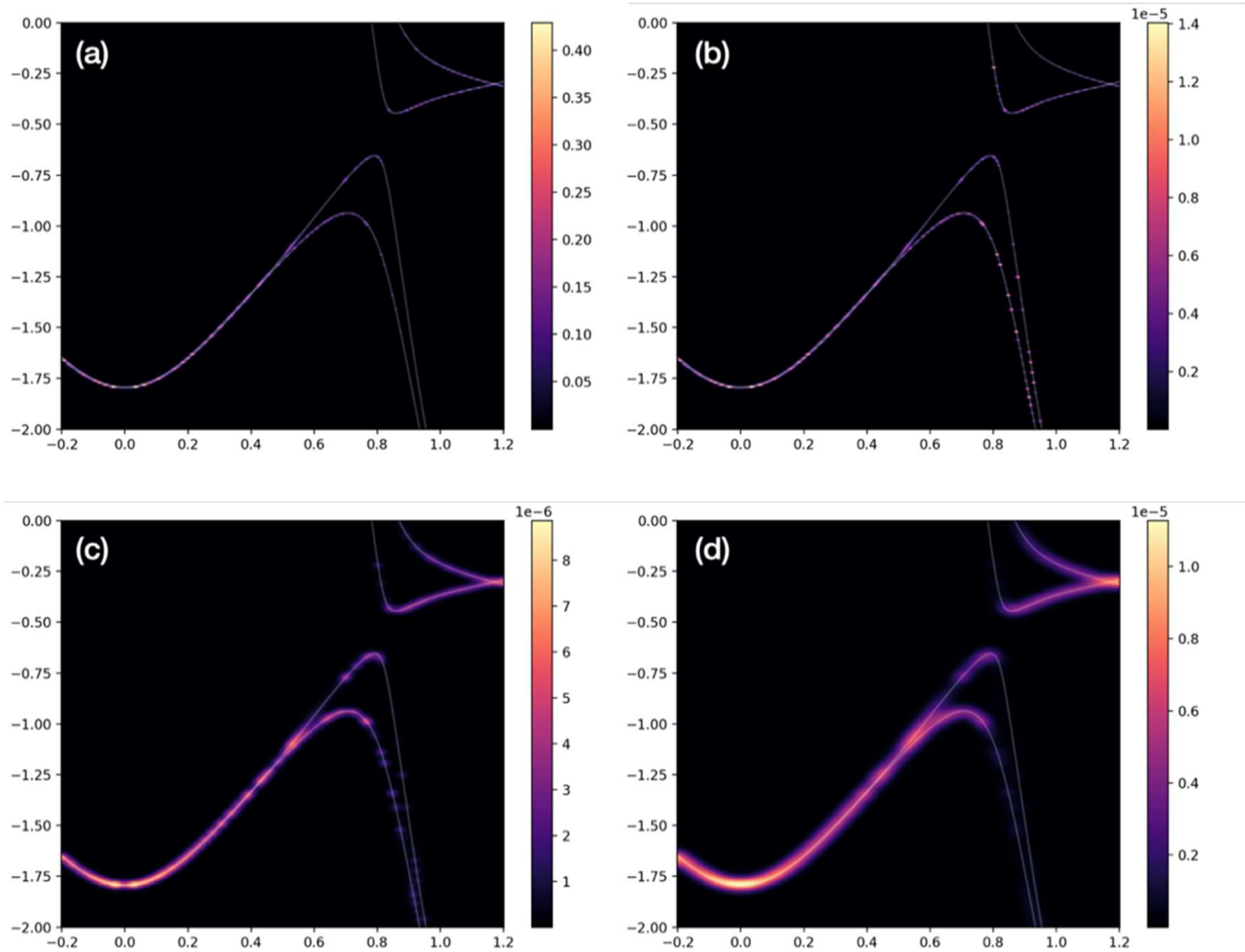

FIG S9: **Matrix elements and finite-resolution effects in ARPES simulation**. The gray solid lines are the plain band structure. The colored and smeared spots show the relative intensity in ARPES simulation. (a) Light polarization along (0, 0, 1) with a smearing of 0.01 for both energy (eV) and momentum ($Å^{-1}$). The $p_x/p_y$ bands are very faint compared to the $p_z$ bands. (b, c, d) Light polarization along (1, 0, 0) with smearing of 0.01, 0.03 and 0.05. The overall brightness for this polarization is 5 orders of magnitude lower than that in (a). In (b) the relative brightness of the $p_x/p_y$ bands is comparable to that of the $p_z$ bands, but it becomes much fainter when the smearing is increased in (c) and (d).

Matrix elements (Figure S9) play a significant role in weakening ARPES signals from the $p_x/p_y$ bands in case (a) where the light polarization is along (0,0,1). If we change the polarization to (1,0,0), the matrix elements from both bands have comparable amplitude, although the absolute values are 5 orders of magnitude smaller than in case (a). This means the matrix elements effect alone cannot explain the band "truncation". In real samples, signals are smeared, and this smearing brings the density of states of the bands into play. After adding the smearing effect, as shown in cases (c) and (d), the signal from the $p_x/p_y$ bands is much weaker than that of the $p_z$ bands. For comparison, the experimental data show a resolution of ~0.1 eV and ~0.05 $Å^{-1}$. Matrix elements and finite-resolution effects[11] together may thus explain the band truncation in the ARPES data.

**Band structures of alternative Pb configurations.**

The band structures of alternative registries with Pb atoms above C atoms or the hollow site exhibit features not observed in ARPES. Figure S10 shows that the band structure of the intermediate registry in the domain boundary, i.e., Pb atoms on top of the C and the hollow site of the uppermost SiC hexagonal lattice, even if periodically maintained under SiC periodicity, is not compatible with ARPES band structure. For the band structure of the hollow site, the spin splitting induced by spin-orbit coupling for the Pb-associated bands along Γ-M around the Fermi level is ~100 meV, much smaller than to the splitting observed in ARPES (~500 meV in Figure S8). Both alternative registries also fail to describe how the band that starts around –1.25 eV at Γ reaches the Fermi level at K. The band structure for the 1×1 Si-top registry, which reflects the interiors of the Frenkel-Kontorova domains, captures both these features.

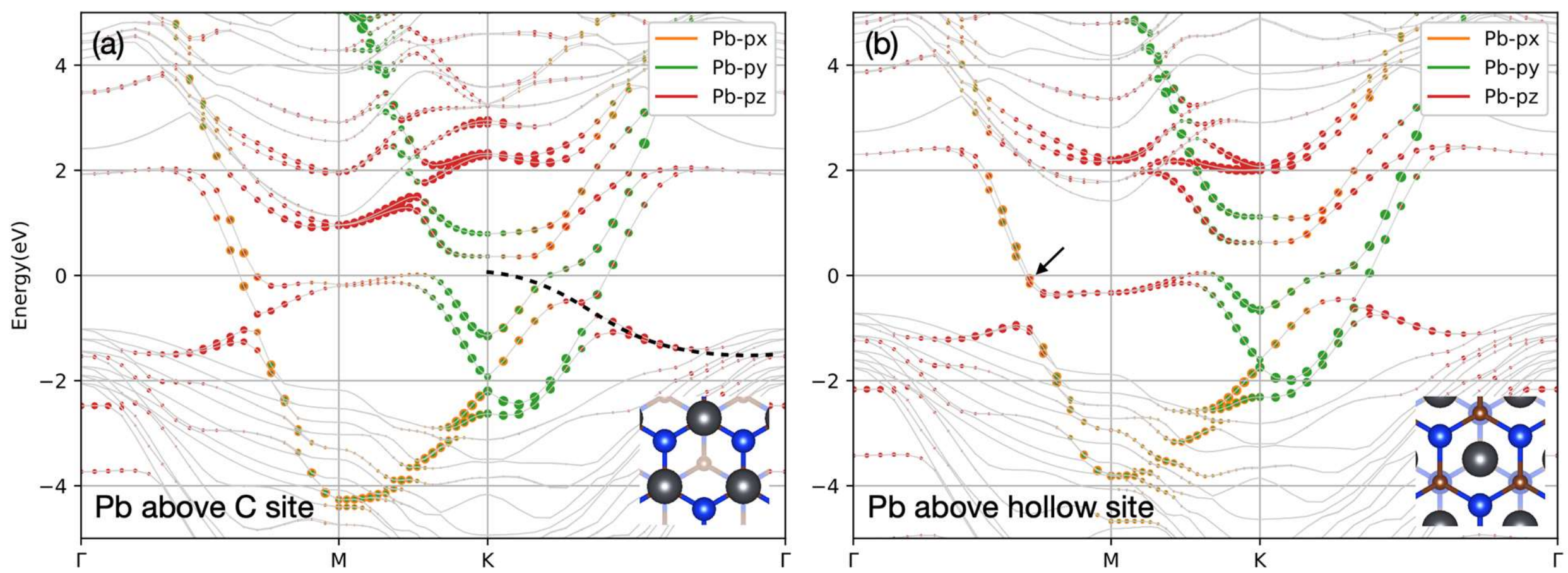

FIG S10: Band structure with spin-orbit coupling of (a) Pb above the C site and (b) Pb above the hollow site of the uppermost SiC layer. The spin splitting at the arrow in (b) is ~100 meV, much smaller than the splitting observed at this position in ARPES. Both band structures do not properly describe the ARPES band around –1.25 eV at Γ that reaches the Fermi level at K, marked by dashed line in (a). The lack of this feature in the simulation of these registries suggests that they occupy at most a small areal fraction of the experimental sample and are not present as extended domains.

B. Matta et al.[12] suggest a $2 \times 2$ SiC-$(\sqrt{3} \times \sqrt{3})R30°$ Pb supercell as a means to match the lattice constants of SiC (0001) and Pb (111). This 30° rotation also allows for weak strain between Pb and graphene in a 7×7 Pb – 10×10 graphene supercell while maintaining the correct relative orientation of graphene to SiC, thus also a possible explanation for the 10×10 Moiré pattern seen in LEED. However, Figure S11 shows that the calculated band structure of the $2 \times 2$ SiC-$(\sqrt{3} \times \sqrt{3})R30°$ Pb system, when unfolded[13,14] to the SiC primitive cell, produces several bands around –1 eV along Γ-to-M that are not visible in ARPES.

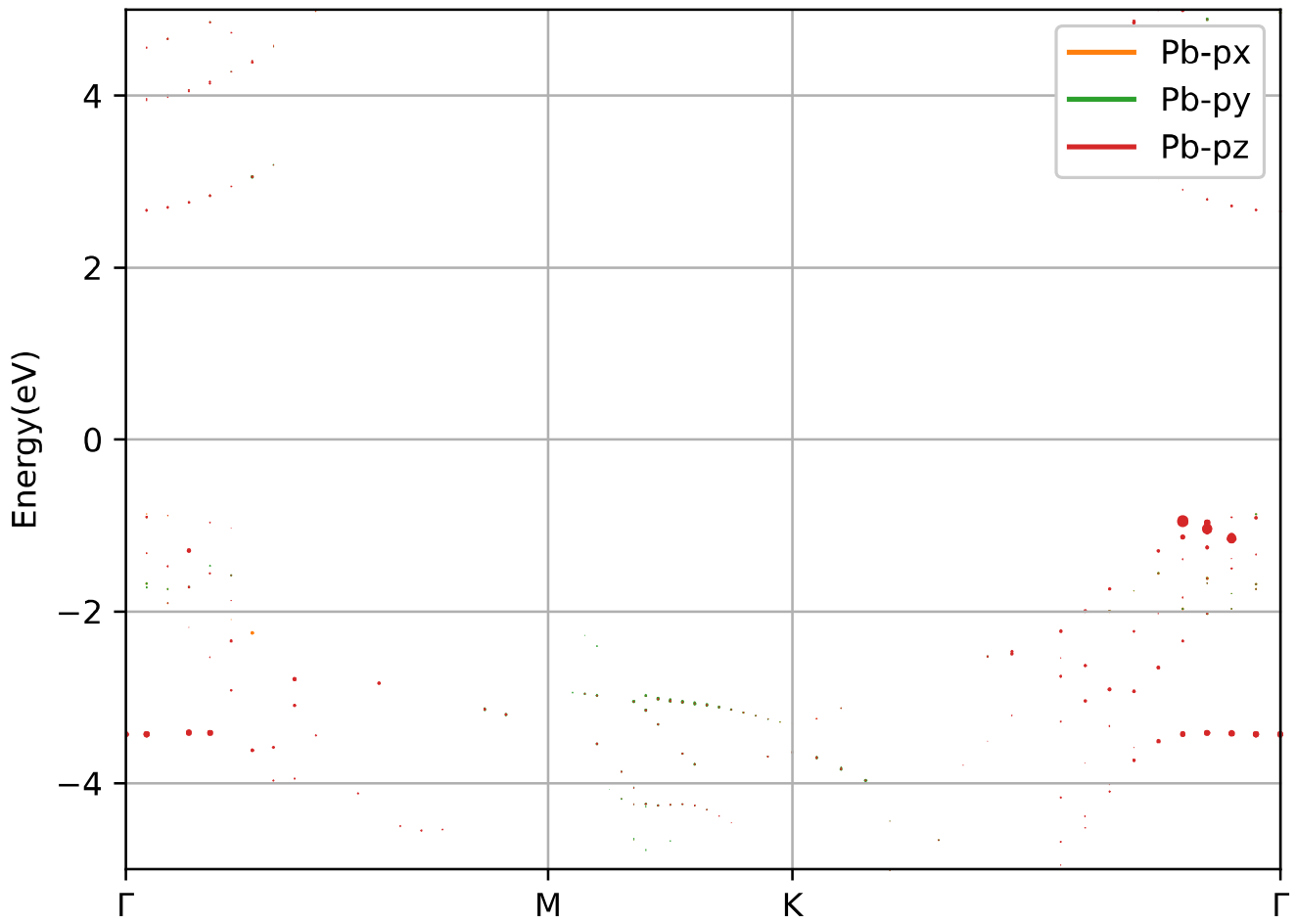

FIG S11: Band structure with spin-orbit coupling of a 2×2 SiC-$\sqrt{3}\times\sqrt{3}$R30° Pb supercell unfolded to the primitive SiC cell. Opacity of dots represents the strength of the spectral weight. Several bands seen around –1 eV along Γ-to-M are not visible in ARPES.

Band structure calculation for 2 layers of Pb on SiC (Figure S12) show two distinct sets of bands of $p_z$ character from Γ to M (and Γ to K) which differ in energy by ~2 eV (i.e. much larger than the Pb monolayer spin-splitting energy scale of 0.5 eV seen in Figure S8). Similarly, we would expect multiple sets of bands of $p_z$ character for thicker Pb intercalation. The ARPES data only show one set of bands in this region, which supports the conclusion that Pb forms a single layer in our experiments.

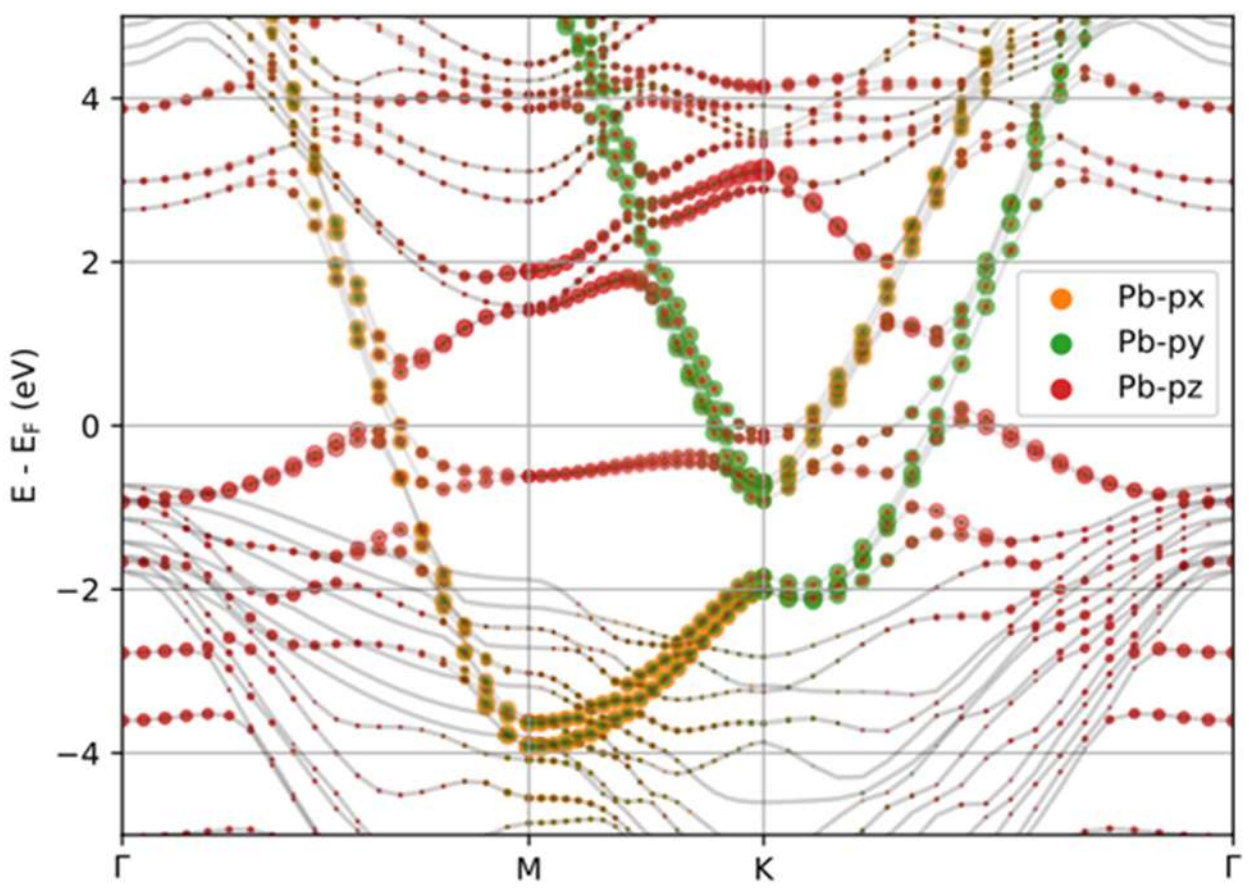

**Phase stability of 1×1 registry.**

We calculated (with spin-orbit coupling) the free energy $E - \mu_{\mathrm{Pb}} N_{\mathrm{Pb}}$ for 1/3, 2/3, and full Pb coverage of the 1×1 registry with two layers of graphene cap as a function of Pb chemical potential $\mu_{\mathrm{Pb}}$. We are referring Gr/SiC with no Pb interclataion to 0 free energy so a phase with positive free energy will not happen during the growth at that $\mu_{\mathrm{Pb}}$. We see that Pb marginally intercalates a full layer into Gr/SiC with a small window before Pb crystallization.

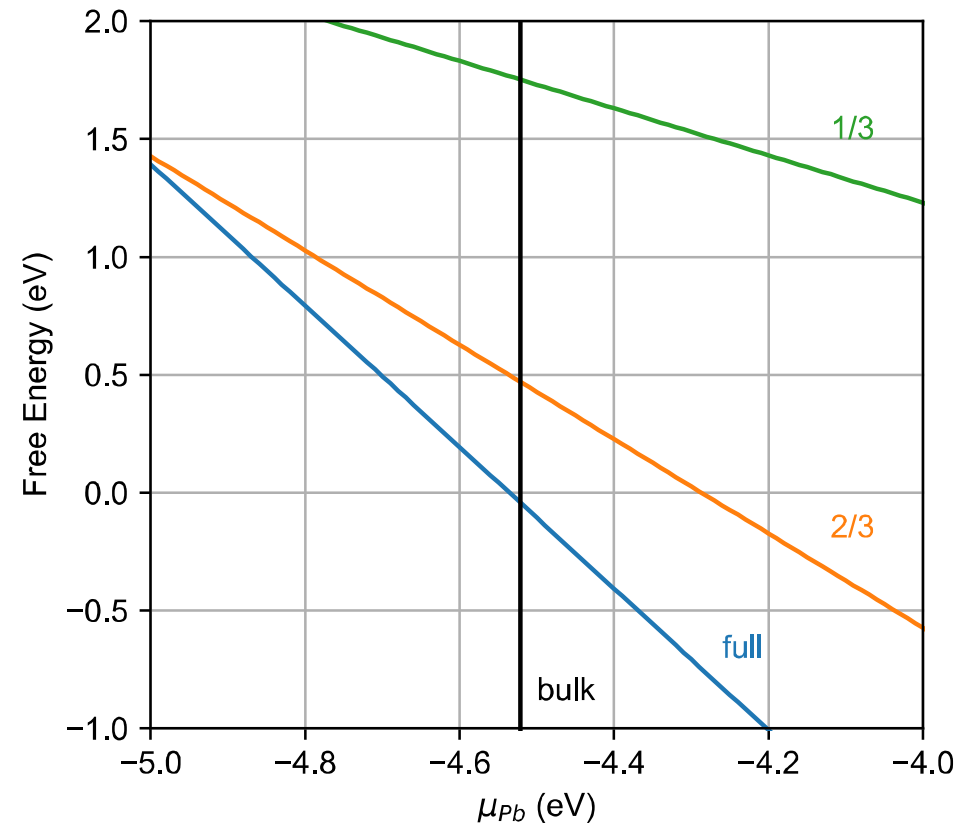

FIG S13: The calculated free energy for different Pb intercalation of 1×1 registry. 0 free energy is referring to no Pb intercalation. "bulk" is referring to the energy of bulk Pb. Only a small window exists allowing Pb intercalation.

## Spin polarization calculations.

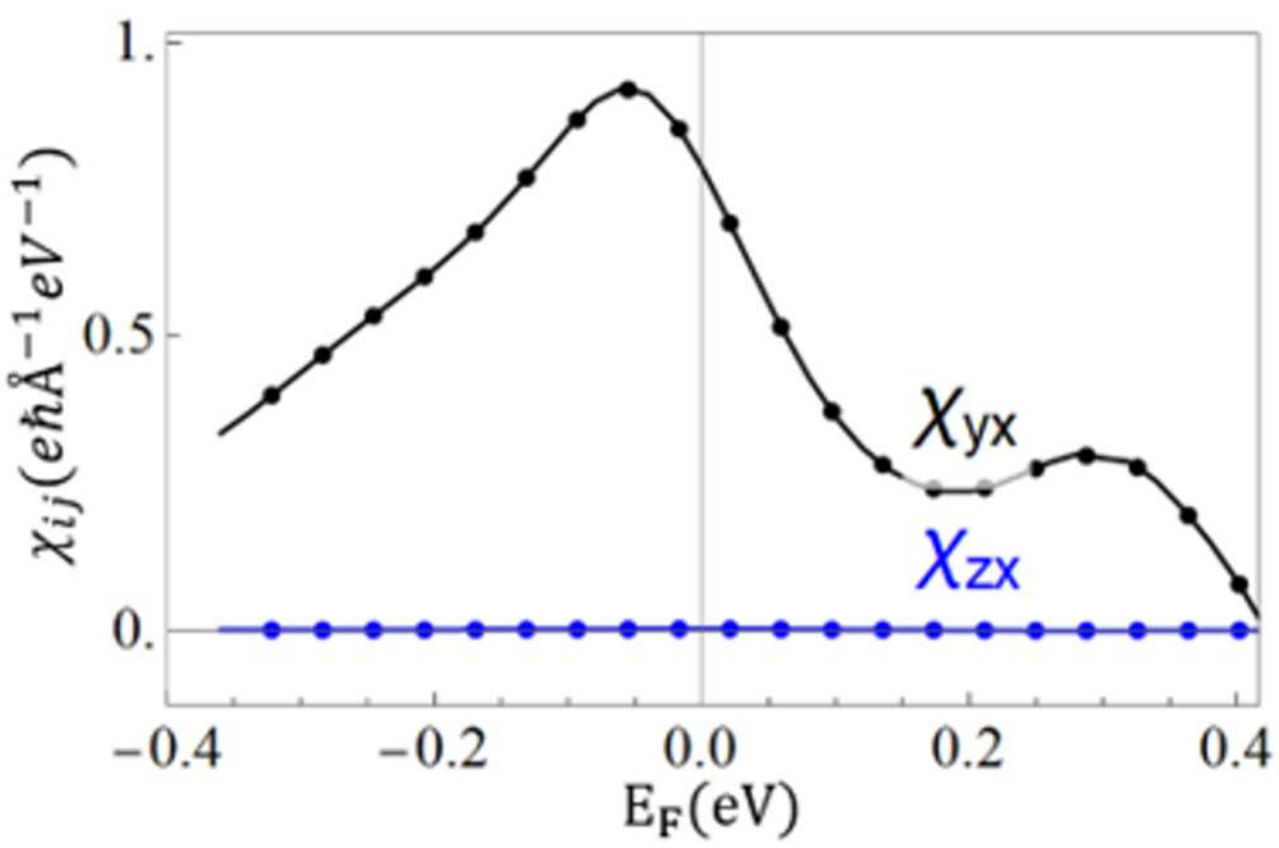

FIG S14: The in-plane CISP response $\chi_{yx}$ and out-of-plane CISP response $\chi_{zx}$.

## XPS fitting parameters.

| | C 1s | | | | | | | | | | | | | | |
|---|---|---|---|---|---|---|---|---|---|---|---|---|---|---|---|
| | sp2 (Gr) LF(0.82,1.5,91,90) | | | SiC/SiC' LF(1,1,255,380,5) | | | Pb-SiC LF(1,1,255,380,5) | | | S1 LA(1.53,243) | | | S2 LA(1.53,243) | | |
| Sample | Position (eV) | FWHM (eV) | Area (%) | Position (eV) | FWHM (eV) | Area (%) | Position (eV) | FWHM (eV) | Area (%) | Position (eV) | FWHM (eV) | Area (%) | Position (eV) | FWHM (eV) | Area (%) |
| Pristine EG | 284.5 | 0.9 | 60 | 283.5 | 0.7 | 29 | | | | 285.1 | 0.7 | 4 | 285.6 | 1 | 7 |
| 2D-Pb_820C | 284.5 | 0.7 | 67 | 283.2 | 0.7 | 7 | 282.8 | 0.7 | 26 | | | | | | |
| 2D-Pb_850C | 284.5 | 0.7 | 71 | 283.1 | 0.8 | 7 | 282.4 | 0.8 | 22 | | | | | | |
| 2D-Pb_880C | 284.5 | 0.7 | 70 | 283.2 | 0.8 | 10 | 282.6 | 0.8 | 20 | | | | | | |
| 2D-Pb_910C | 284.5 | 0.7 | 69 | 283.1 | 0.9 | 10 | 282.5 | 0.9 | 21 | | | | | | |
| 2D-Pb_925C | 284.5 | 0.7 | 68 | 283.1 | 0.8 | 6 | 282.4 | 0.8 | 26 | | | | | | |
| 2D-Pb_935C | 284.5 | 0.7 | 69 | 283.2 | 0.8 | 5 | 282.4 | 0.8 | 24 | | | | | | |
| 2D-Pb_940C | 284.5 | 0.7 | 73 | 283.2 | 0.8 | 6 | 282.7 | 0.8 | 21 | | | | | | |
| hEG | 284.5 | 0.7 | 51 | 282.6 | 0.8 | 49 | | | | | | | | | |

| | Si 2p | | | | | | | | | | | |
|---|---|---|---|---|---|---|---|---|---|---|---|---|
| | SiC 2p3/2 LF(1,1,255,380,5) | | | SiC 2p1/2 LF(1,1,255,380,5) | | | Pb-Si 2p3/2 LF(1,1,255,380,5) | | | Pb-Si 2p1/2 LF(1,1,255,380,5) | | |
| Sample | Position (eV) | FWHM (eV) | Area (%) | Position (eV) | FWHM (eV) | Area (%) | Position (eV) | FWHM (eV) | Area (%) | Position (eV) | FWHM (eV) | Area (%) |
| Pristine EG | 101.3 | 0.9 | 67 | 101.9 | 0.9 | 33 | | | | | | |
| 2D-Pb_820C | 100.9 | 0.7 | 19 | 101.5 | 0.7 | 9 | 100.4 | 0.7 | 46 | 101.0 | 0.7 | 23 |
| 2D-Pb_850C | 100.8 | 0.9 | 17 | 101.5 | 0.9 | 8 | 100.1 | 0.9 | 48 | 100.7 | 0.9 | 24 |
| 2D-Pb_880C | 100.8 | 1.0 | 24 | 101.5 | 1.0 | 12 | 100.1 | 1.0 | 39 | 100.7 | 1.0 | 19 |
| 2D-Pb_910C | 100.8 | 1.0 | 23 | 101.4 | 1.0 | 11 | 100.1 | 1.0 | 40 | 100.8 | 1.0 | 20 |
| 2D-Pb_925C | 100.7 | 1.0 | 13 | 101.3 | 1.0 | 6 | 100.0 | 1.0 | 52 | 100.6 | 1.0 | 26 |
| 2D-Pb_935C | 100.7 | 1.0 | 13 | 101.4 | 1.0 | 6 | 100.0 | 1.0 | 52 | 100.6 | 1.0 | 26 |
| 2D-Pb_940C | 100.8 | 0.9 | 14 | 101.4 | 0.9 | 6 | 100.3 | 0.9 | 51 | 100.9 | 0.9 | 25 |
| hEG | 100.3 | 0.9 | 67 | 100.9 | 0.9 | 33 | | | | | | |

**Table 1:** Fitting parameters[7] for the C 1s and Si 2p high-resolution spectra. Samples are all charge corrected (sp2 at 284.5 eV). LA denotes a Lorentzian centered at $E_0$ of the form:

$$LA(a,b,m) \ or \ LF(a,b,w,m) = \begin{cases} [L(E)]^a \ for \ E \leq E_0 \\ [L(E)]^b \ for \ E > E_0 \end{cases}$$

Where:

$$L(E) = \frac{1}{1 + 4(\frac{E - E_0}{FWHM})^2}$$

Convoluted with a Gaussian of width characteristic $m$ to form a Voigt line shape. A damping function is applied ($w$) to force tails to the baseline within the region bounds.